\documentclass[12pt,letterpaper]{article}

\usepackage{subcaption}
\usepackage{amsmath,amssymb,calc}
\usepackage{tikz}
\usepackage{graphicx}
\usepackage{cite}
\usepackage[T1]{fontenc}
\usepackage{float}
\usepackage{latexsym}
\usepackage{amsmath,amstext,amssymb,amsfonts,
amscd,bm,array,multirow,amsbsy,mathrsfs,amsthm,calc}
\usepackage{t1enc}
\usepackage{indentfirst}
\usepackage{pb-diagram}
\usepackage{graphicx}
\usepackage{enumerate}
\usepackage{color}
\usepackage[all]{xy}
\usepackage{url}
\usepackage{upgreek}

\newcommand{\eqdef}{\stackrel{\mathrm{def.}}{=}}

\newcommand{\bea}{\begin{eqnarray}}
\newcommand{\eea}{\end{eqnarray}}

\def\cM{\mathcal{M}}

\newcommand{\pd}{\partial}

\def\id{\protect{{1 \kern-.28em {\rm l}}}}

\def\cG{{\cal G}}

\def\pd{\partial}

\def\dd{\mathrm{d}}

\def\const{\mathrm{const}}

\newcommand{\be}{\begin{equation*}}
\newcommand{\ee}{\end{equation*}}
\newcommand{\ben}{\begin{equation}}
\newcommand{\een}{\end{equation}}
\newcommand{\beqa}{\begin{eqnarray*}}
\newcommand{\eeqa}{\end{eqnarray*}}
\newcommand{\beqan}{\begin{eqnarray}}
\newcommand{\eeqan}{\end{eqnarray}}
\newcommand{\nn}{\nonumber}

\def\id{\protect{{1 \kern-.28em {\rm l}}}}

\begin{document}

\begin{titlepage}
\begin{center}
\hfill \\
\vspace{1cm}
{\Large {\bf Dynamical consistency conditions for\\ rapid turn inflation\\[3mm] }}

\vskip 1.5cm
{\bf Lilia Anguelova${}^a$\footnote{anguelova@inrne.bas.bg}, Calin Iuliu Lazaroiu${}^b$\footnote{lcalin@theory.nipne.ro}\\
\vskip 0.5cm {\it ${}^a$ Institute for Nuclear Research and Nuclear Energy,\\ Bulgarian Academy of Sciences, Sofia, Bulgaria\\ ${}^b$ Horia Hulubei National Institute for Physics and Nuclear\\ Engineering (IFIN-HH), Bucharest-Magurele, Romania}}

\vskip 6mm

\end{center}

\vskip .1in
\vspace{0.5cm}

\begin{center} {\bf Abstract}\end{center}

\vspace{-1cm}

\begin{quotation}\noindent

We derive consistency conditions for sustained slow roll and rapid
turn inflation in two-field cosmological models with oriented scalar
field space, which imply that inflationary models with 
field-space trajectories of this type are non-generic. In particular, we show that
third order adiabatic slow roll, together with large and slowly varying turn rate,
requires the scalar potential of the model to satisfy a certain
nonlinear second order PDE, whose coefficients depend on the scalar
field metric. We also derive consistency conditions for slow roll 
inflationary solutions in the so called ``rapid turn attractor'' approximation, 
as well as study the consistency conditions for circular rapid turn trajectories 
with slow roll in two-field models with rotationally invariant field space metric. 
Finally, we argue that the rapid turn regime tends to have a natural exit after 
a limited number of e-folds.
\end{quotation}

\end{titlepage}

\eject

\tableofcontents

\section{Introduction}

Modern observations have established -- to a very good degree of
accuracy -- that the present day universe is homogeneous and isotropic
on large scales. This is naturally explained if one assumes that the
early universe underwent a period of accelerated expansion called
inflation. This idea can be realized in models where the inflationary
expansion is driven by the potential energy of a number of real scalar
fields called inflatons. The most studied models of this type contain
a single scalar field.  However, recent arguments related to quantum
gravity suggest that it is more natural, or may even be necessary
\cite{GK,OPSV,AP,BPR} to have more than one inflaton. This has generated
renewed interest in multi-field cosmological models, which had previously
attracted only limited attention. 

Multifield cosmological models have richer phenomenology than single
field models since they allow for solutions of the equations of motion
whose field-space trajectories are not (reparameterized)
geodesics. Such trajectories are characterized by a non-zero {\em turn
rate}. In the past it was thought that phenomenological viability
requires small turn rate, by analogy with the slow roll approximation
used in the single-field case. This assumption leads to the celebrated
slow-roll slow-turn (SRST) approximation of \cite{PT1,PT2}. However,
in recent years it was understood that rapid turn trajectories can
also be (linearly) perturbatively stable \cite{CAP,BFM} and of
phenomenological interest. For instance, a brief rapid turn during
slow-roll inflation can induce primordial black hole generation
\cite{PSZ,FRPRW,LA,LA2}; moreover, trajectories with large and constant
turn rate can correspond to solutions behaving as dark energy
\cite{ASSV,ADGW}.  There is also a variety of proposals for
full-fledged rapid-turn inflation models, relying on large turn rates
during the entire inflationary period
\cite{AB,SM,TB,BM,GSRPR,CRS,ACIPWW,APR}.

Finding inflationary solutions in multifield models is much harder
than in the single-field case, because the background field equations
form a complicated coupled system of nonlinear ODEs. Thus usually such
models are either studied numerically or solved only
approximately.\footnote{A notable exception is provided by models
with hidden symmetry, which greatly facilitates the search for exact
solutions \cite{ZGZ,PT,ABL,ABL2,Hesse}.} Mathematically, this complicated
coupled system is encoded by the so-called {\em cosmological
equation}, a nonlinear second order geometric ODE defined on the
scalar field space of the model. The latter is a connected paracompact
manifold, usually called the {\em scalar manifold}. In turn, the
cosmological equation is equivalent with a dissipative geometric
dynamical system defined on the tangent bundle of that
manifold. Little is known in general about this dynamical system, in
particular because the scalar manifold need not be simply-connected
and -- more importantly -- because this manifold is non-compact in
most applications of physical interest and hence cosmological
trajectories can ``escape to infinity''. The resulting dynamics can be
surprisingly involved\footnote{It is sometimes claimed that the
complexity of this dynamics could be ignored, because in
``phenomenologically relevant
models'' one should ``expect that'' all directions orthogonal to the
physically relevant scalar field trajectory are heavy and hence can
be integrated out, thus reducing the analysis to that of a
single-field model. This argument is incorrect for a number of
reasons.  First, current phenomenological data does {\em not} rule out
multifield dynamics. Second, such a reduction to a one field model
(even when possible) relies on knowledge of an appropriate
cosmological trajectory, which itself must first be found by analyzing
the dynamics of the multifield model.} and hard to analyze even by
numerical methods (see \cite{genalpha,elem,modular} for some
nontrivial examples in two-field models), though a conceptual approach
to some aspects of that dynamics was recently proposed in
\cite{ren,grad}.

A common approach to looking for cosmological trajectories with
desirable properties is to first simplify the equations of motion by
imposing various approximations (such as slow-roll to a certain order,
rapid-turn and/or other conditions). This leads to an approximate
system of equations, obtained by neglecting certain terms in the
original ODEs. Then one attempts to solve the approximate system
numerically or analytically. However, there is apriori no guarantee
that a solution of the approximate system is a good approximant of a
solution of the exact system for a sufficiently long period of
time. In general, this will be the case only if the data which
parameterizes the exact system (namely the scalar field metric and
scalar potential of the model) satisfies appropriate {\em consistency
conditions}. Despite being of fundamental conceptual importance, such
consistency conditions have so far not been studied systematically
in the literature.

In this paper, we investigate the problem of consistency conditions in
two-field models with orientable scalar manifold for several commonly
used approximations. First, we consider third order slow-roll
trajectories with large but slowly varying turn rate. In this case, we
show that compatibility with the equations of motion requires that the
scalar potential satisfies a certain nonlinear second order PDE whose
coefficients depend on the scalar field metric. This gives a
nontrivial and previously unknown consistency condition that must be
satisfied in the field-space regions where one can expect to find
sustained rapid-turn trajectories allowing for slow roll
inflation. Therefore, inflationary solutions of this type are not easy
to find, implying that two-field models with such families of
cosmological trajectories are non-generic.  In particular, this shows
that the difficulty in finding such models which was noticed in
\cite{ACPRZ} is not related to supergravity, but arises on a more
basic level.

We also discuss the case of rotationally invariant scalar field-space
metrics. In that case, it is common to consider field space
trajectories which are nearly circular as candidates for sustained
rapid turn inflation. Imposing the first and second order
slow roll conditions in this context leads to a certain consistency
condition for compatibility with the equations of motion. This is
again a PDE for the scalar potential with coefficients depending on
the field space metric, which does not seem to have been widely noticed 
in the literature. We consider its implications for important examples 
in previous work.

Finally, we study in detail the consistency conditions for the
approximation of \cite{TB}, which subsumes many prominent rapid turn
models of inflation. We show that this approximation is a special case
of rapid turn with slow roll, instead of being equivalent to it. We
then derive conditions for compatibility of this approximation with
the equations of motion. Once again, these constrain the
scalar potential and field space metric.

Throughout the paper, we assume that the field space (a.k.a. scalar
manifold) of the two-field model is an {\em oriented} and connected
paracompact surface. In our considerations, a crucial role is played
by a fixed oriented frame (called the {\em adapted frame}) of vector
fields defined on this surface which is determined by the scalar
potential and field-space metric, instead of the moving oriented
Frenet frame determined by the field space trajectory. The two frames
are related to each other through a time-dependent rotation whose
time-dependent angle we call the {\em characteristic angle}. We
conclude our investigations by studying the time evolution of this
angle. We show that, generically, this angle tends rather fast to the
value $\pi \mod 2\pi$, which means that the tangent vector of the
inflationary trajectory aligns with minus the gradient of the
potential. This implies that the rapid turn regime has a natural exit
in the generic case. Our results also suggest that it is difficult to
sustain this regime for a prolonged period.

The paper is organized as follows. Section \ref{sec:intro} recalls
some basic facts about two-field cosmological models and introduces
various parameters which will be used later on. Section \ref{sec:cons}
discusses the consistency condition for sustained rapid turn
trajectories with third order slow roll, which results from careful
analysis of the compatibility of the corresponding approximations with
the equations of motion. Section \ref{sec:rot} discusses the
consistency condition for circular trajectories in
rotationally-invariant models, as well as some implications for
previous work on such inflationary trajectories. Section
\ref{sec:Bjorkmo} discusses the approximation of \cite{TB}, showing
how it differs from rapid turn with second order slow roll and
extracts the relevant consistency conditions. Section
\ref{sec:charangle} studies the time evolution of the characteristic
angle, while Section \ref{sec:conclusions} presents our conclusions.

\section{Two-field cosmological models}
\setcounter{equation}{0}
\label{sec:intro}

The action for $n$ real scalar fields $\varphi^I (x^{\mu})$
minimally coupled to four-dimensional gravity is:
\ben
\label{Action_gen}
S = \int d^4x \sqrt{-\det g} \left[ \frac{R(g)}{2} - \frac{1}{2} G_{IJ} (\{\varphi^I\}) \pd_{\mu} \varphi^I \pd^{\mu} \varphi^J - V (\{ \varphi^I \}) \right] \,\,\, ,
\een
where we took $M_{\mathrm{Pl}}=1$. Here $g_{\mu \nu}$ is the spacetime
metric (which we take to have ``mostly plus'' signature) and $R(g)$ is
its scalar curvature. The indices $\mu,\nu$ run from from $0$ to $3$
and $G_{IJ}$ is the metric on the scalar manifold (a.k.a. ``scalar
field space'') $\cM$, which is parameterized by the scalars
$\{\varphi^I\}$ with $I,J = 1,...,n$. The scalar potential $V$ is a
smooth real-valued function defined on $\cM$, which for simplicity we
take to be strictly positive. The standard cosmological Ansatze for
the background metric and scalar fields are:
\ben
\label{metric_g}
\dd s^2_g = -\dd t^2 + a^2(t) \dd \vec{x}^2 \qquad , \qquad \varphi^I = \varphi^I (t) \quad ,
\een 
where $a(t)>0$ is the scale factor. As usual, the definition of the
Hubble parameter is:
\ben
\label{Hp}
H (t) = \frac{\dot{a}}{a} \,\,\, ,
\een
where the dot denotes derivation with respect to $t$.

With the Ansatze above, the equations of motion for the scalars reduce to:
\ben
\label{EoM_sc}
D_t \dot{\varphi}^I + 3 H \dot{\varphi}^I + G^{IJ} V_J = 0 \quad ,
\een
where $V_J \eqdef \pd_J V$ with $\pd_J \eqdef \pd_{\varphi^J}$ and we defined:
\be
D_t A^I \eqdef \dot{\varphi}^J \,\nabla_J A^I = \dot{A}^I + \Gamma^I_{JK}(\varphi) \dot{\varphi}^J A^K
\ee
for any vector field $A^I$ defined on the scalar manifold, where
$\Gamma^I_{JK}$ are the Christoffel symbols of $G_{IJ}$. The Einstein
equations can be written as:
\ben
\label{EinstEqs}
G_{IJ} \dot{\varphi}^I \dot{\varphi}^J = - 2 \dot{H}  \qquad  ,  \qquad  3 H^2 + \dot{H} = V \qquad .
\een

In this paper we focus on two-field models, i.e. we take $n=2$. In
this case, the scalar manifold $(\cM,\cG)$ is a (generally non-compact) Riemann
surface, which we assume to be oriented. To define various
characteristics of an inflationary solution, one
introduces a frame of tangent vectors to the field space. A widely used
choice is the positive Frenet frame, which consists of the tangent
vector $T(t)$ and normal vector $N(t)$ to the trajectory
$\varphi(t)=(\varphi^1(t),\varphi^2(t))\in \cM$ of a solution of
\eqref{EoM_sc}, where the normal vector is chosen such that $(T(t),N(t))$
forms a positively-oriented basis of the tangent space of the scalar manifold $\cM$
at the point $\varphi(t)$. Let $\sigma$
be an increasing proper length parameter along the solution curve
$\varphi$. This parameter is determined up to a constant translation
and satisfies:
\be
\dot{\sigma} = \sqrt{G_{IJ} \dot{\varphi}^I \dot{\varphi}^J}~~.
\ee
Then the vectors $T$ and $N$ are given by:
\be
T^I = \frac{\dd\varphi^I}{\dd \sigma}=\frac{\dot{\varphi}^I}{\dot{\sigma}} 
\ee
and 
\ben
\label{N_def}
N_I = (\det G)^{1/2} \epsilon_{IJ} T^J \,\,\, .
\een
One has:
\ben
\label{Orthon}
N_I T^I = 0 \qquad , \qquad N_I N^I = 1 \qquad , \qquad T_I T^I = 1 \quad ,
\een
where $T_I = G_{IJ} T^J$ , $N^I = G^{IJ} N_J$\,. 

\subsection{Characteristics of an inflationary solution} \label{TwoF}

\noindent Consider the {\em opposite relative acceleration vector}:
\ben
\label{SR_par}
\eta^I \eqdef - \frac{1}{H \dot{\sigma}} D_t \dot{\varphi}^I \,\,\, .
\een
Expanding $\eta$ in the orthonormal basis $(T,N)$ gives:
\be
\eta^I = \eta_{\parallel} T^I + \eta_{\perp} N^I \,\,\, ,
\ee
with:
\ben
\label{eta_PP}
\eta_{\parallel} = - \frac{\ddot{\sigma}}{H \dot{\sigma}} \qquad {\rm and} \qquad \eta_{\perp} = \frac{\Omega}{H} \,\,\, ,
\een
where we defined the {\em signed turn rate} of the trajectory $\varphi$ by:
\ben
\label{Om_1}
\Omega \eqdef - N_I D_t T^I \,\,\, .
\een
The quantity $\eta_\parallel$ is called the {\em second slow
  roll parameter} of $\varphi$, while $\eta_\perp$
(which is sometimes denoted by $\omega$) is called the {\em first turn
  parameter} (or signed reduced turn rate). We will see in a moment that
$\eta_\parallel$ is the second slow roll parameter of a certain
one-field model determined by the trajectory $\varphi$.
The {\em second turn parameter} is defined through:
\ben \label{nu_def}
\nu\eqdef \frac{\dot{\eta}_{\perp}}{H \eta_{\perp}}~~.
\een
It is a measure for the rate of change per e-fold of the dimensionless 
turn parameter $\eta_{\perp}$\,.

Projecting the scalar field equations (\ref{EoM_sc}) along $T^I$ gives
the {\em adiabatic equation}:
\ben
\label{EoM_sigma}
\ddot{\sigma} + 3 H \dot{\sigma} + V_T = 0 \,\,\, ,
\een
where $V_T \eqdef T^I V_I$\,. Projecting (\ref{EoM_sc}) along
$N_I$\, gives the {\em entropic equation}:
\ben
\label{EoM_N}
N_I D_t T^I = - \frac{V_N}{\dot{\sigma}} \,\,\, ,
\een
where $V_N\eqdef N^I V_I$. Using the definition \eqref{Om_1}, the
entropic equation (\ref{EoM_N}) reads:
\ben
\label{Om_2}
\Omega = \frac{V_N}{\dot{\sigma}} \,\,\, .
\een

Let $V_\varphi(\sigma)\eqdef V(\varphi(\sigma))$. Then:
\be
V_T=\frac{\dd V_\varphi}{\dd \sigma}=V'_\varphi(\sigma)~~,
\ee
where the prime denotes derivation with respect to $\sigma$. Hence
the adiabatic equation \eqref{EoM_sigma} reads:
\be
\ddot{\sigma} + 3 H \dot{\sigma} + V_\varphi'(\sigma) = 0 \,\,\, ,
\ee
which shows that $\sigma$ obeys the equation of motion of a
one-field model with scalar potential $V_\varphi$. Since for an expanding universe 
we have $H>0$\,, the Hubble parameter can be expressed from \eqref{EinstEqs} as:
\be
H(\sigma ,\dot{\sigma})=\frac{1}{\sqrt{6}}\sqrt{\dot{\sigma}^2+2V(\sigma)}~~,
\ee
which is the usual formula for a single-field model.

The {\em adiabatic Hubble slow roll parameters} of the trajectory
$\varphi$ are defined as the usual Hubble slow roll-parameters
\cite{LPB} of the adiabatic one-field model defined above. The first
three of these are:
\ben \label{3rdOsr_par}
\varepsilon\eqdef-\frac{\dot{H}}{H^2}\qquad , \qquad \eta_\parallel= -\frac{\ddot{\sigma}}{H\dot{\sigma}}\qquad , \qquad
\xi\eqdef \frac{\dddot{\sigma}}{H^2\dot{\sigma}}~~,
\een
where for $\xi$ we use a different sign convention than \cite{LPB}.
Since all results of \cite{LPB} apply to the adiabatic one field
models, it follows that the following relations hold on-shell:
\beqa
&& \epsilon_T=\varepsilon\left(\frac{3-\eta_\parallel}{3-\varepsilon}\right)^2\nn\\
&& \eta_\parallel=\varepsilon -\frac{\varepsilon'}{\sqrt{2\varepsilon}}\\
&& \eta_T=\sqrt{2\varepsilon} \frac{\eta'_\parallel}{3-\varepsilon}+\frac{3-\eta_\parallel}{3-\varepsilon}(\varepsilon+\eta_\parallel)\nn~~,
\eeqa
where the adiabatic first and second {\em potential slow roll parameters} are defined through:
\be
\epsilon_T\eqdef \frac{1}{2}\left(\frac{V_T}{V}\right)^2~~,~~\eta_T\eqdef \frac{V''_\varphi}{V}~~.
\ee

\subsection{Slow roll and rapid turn regimes}

\noindent The first, second and third slow roll regimes are defined
respectively by the conditions $\varepsilon\ll 1$, $|\eta_\parallel|\ll
1$ and $|\xi|\ll 1$.  The second {\em order} slow roll regime is
defined by the two conditions $\varepsilon\ll 1$ and $|\eta_\parallel|\ll
1$ taken together, while the third {\em order} slow roll regime is
defined by the three conditions $\varepsilon\ll 1$, $|\eta_\parallel|\ll
1 $ and $|\xi|\ll 1$ considered together.

The {\em first order rapid turn regime} is defined by the condition:
\ben
\label{Large_eta_perp}
\eta_{\perp}^2 \gg 1 \,\,\, ,
\een
while the condition $|\nu|\ll 1$ defines the {\em second order rapid turn
  regime}. A trajectory has {\em sustained rapid turn}
if both of the conditions $\eta_\perp^2\ll 1$ and $|\nu|\ll 1$ are
satisfied.

In the first slow roll regime $\varepsilon\ll 1$, the second equation
in (\ref{EinstEqs}) becomes:
\ben \label{H_V_sr}
3H^2 \approx V \,\,\, .
\een
Together with (\ref{Om_2}), this gives:
\ben
\label{Eta_p}
\eta_{\perp}^2 = \frac{\Omega^2}{H^2} \approx \frac{3 V_N^2}{\dot{\sigma}^2 V} \,\,\, .
\een

\subsection{Some other useful parameters}

\noindent For our purpose, it will be useful to consider two other
parameters, which are related to the familiar ones reviewed above.
The {\em first IR parameter} is the ratio of the kinetic and potential
energies of the scalars:
\ben
\label{kappa_def}
\kappa \eqdef \frac{G_{IJ} \dot{\varphi}^I \dot{\varphi}^J }{2V} = \frac{\dot{\sigma}^2}{2V} \,\,\, .
\een
This parameter plays an important role in the IR approximation
discussed in \cite{ren} and \cite{grad}. On solutions of the equations of
motion, the first slow roll parameter can be written as:
\be
\varepsilon = \frac{3 \dot{\sigma}^2}{\dot{\sigma}^2+2V} \,\,\, .
\ee
Hence the on-shell value of the first slow roll parameter is
related to the first IR parameter through:
\ben
\label{ep_kappa}
\varepsilon = \frac{3 \kappa}{1 + \kappa} \,\,\, .
\een
This relation shows that the first slow roll condition $\varepsilon \ll 1$ is equivalent
with $\kappa \ll 1$\,, whereas $\varepsilon = 1$ corresponds to
$\kappa = \frac{1}{2}$\,.  In particular, the condition for
inflation $\varepsilon < 1$ is equivalent with $\kappa<\frac{1}{2}$.

We also define the {\em conservative parameter}:
\ben
\label{c_def}
c \eqdef \frac{H \dot{\sigma}}{\sqrt{G^{IJ} V_I V_J}} \,\,\, ,
\een
which is small iff the friction term can be neglected relative to the
gradient term in the equation of motion for the scalar fields. When
$c\ll 1$, the motion of $\varphi$ is approximately conservative in the
sense that the total energy
$\frac{1}{2}G_{IJ}\dot{\varphi}^I\dot{\varphi}^J$ of the scalar fields
is approximately conserved; in this limit, the equations of motion for
the scalars reduce to:
\be
\nabla_t \varphi^I+G^{IJ}\pd_J V\approx 0~~,
\ee
which describe the motion of a particle in the Riemannian manifold $(\cM,\cG)$
under the influence of the potential $V$. We will see below that, in the second slow
roll regime $|\eta_\parallel|\ll 1$, the rapid turn condition
(\ref{Large_eta_perp}) is equivalent with the {\em conservative
  condition}:
\be
c^2 \ll 1 \,\,\, .
\ee

\section{The consistency condition for sustained slow roll with rapid turn}
\setcounter{equation}{0}
\label{sec:cons}

In this section, we derive a constraint relating the scalar potential
$V$ and the field-space metric $G_{IJ}$\, which is necessary for
existence of rapid turn solutions ($\eta_\perp^2\gg 1$) that satisfy
the {\em third order} slow roll conditions $\varepsilon\ll 1$,
$|\eta_\parallel|\ll 1$ and $|\xi|\ll 1$\, as well as the second order 
condition $|\nu|\ll 1$ for slowly-varying turn rate. The last
condition is usually imposed to ensure the longevity of the rapid-turn
slow-roll regime (so that it could produce about 50-60 e-folds of
inflation). The constraint we derive does not require any extraneous
assumptions about the potential, the field space metric or the shape
of the field-space trajectory.

\subsection{The adapted frame}

\noindent We will use a globally-defined frame $(n,\tau)$ (which we
call {\em the adapted frame}) of the scalar manifold
determined by $V$ and $G_{IJ}$. Namely, we take $n$ to be the unit
vector field along the gradient of the potential:
\ben
\label{ndef}
n \eqdef \frac{\nabla V}{\sqrt{G^{IJ} V_I V_J}} \quad ,\,\,{\rm with\,\,components} \quad n^K = \frac{G^{KL} V_L}{\sqrt{G^{IJ} V_I V_J}} \,\,\, .
\een
The unit vector $\tau$ is orthogonal to $n$ and chosen such
that the basis $(n,\tau)$ is positively oriented:
\ben \label{taudef}
\tau_I = (\det G)^{1/2} \epsilon_{IJ} n^J \,\,\, . 
\een
We have $\tau^I \eqdef G^{IJ} \tau_J$ as usual. The relation between the two bases is given by:
\beqan
\label{BasesRel}
T &=& ~\,\cos \theta_{\varphi} \,\, n + \sin \theta_{\varphi} \,\, \tau \,\,\, , \nn \\
N &=& -\!\sin \theta_{\varphi} \,\, n + \cos \theta_{\varphi} \,\, \tau \,\,\, ,
\eeqan
where $\theta_{\varphi}\in (-\pi,\pi]$ is the {\em characteristic angle} of $\varphi$, which is
defined as the angle of the rotation that takes the oriented basis $(n,\tau)$ to the oriented
basis $(T,N)$ at the point of interest on the cosmological trajectory. 

\subsection{Expressing $\eta_\parallel$ and $\eta_\perp$ in terms of $\theta_\varphi$ and $c$}

\noindent Relation \eqref{eta_PP} and the adiabatic equation (\ref{EoM_sigma}) give:
\ben
\label{eta_par_c}
\eta_{\parallel} = 3 + \frac{V_{\sigma}}{H \dot{\sigma}} = 3 + \frac{T^I G_{IJ} n^J}{H \dot{\sigma}} \sqrt{G^{KL}V_K V_L} = 3 + \frac{\cos \theta_{\varphi}}{c} \,\,\, ,
\een
where the conservative parameter $c$ was defined in (\ref{c_def}). Notice that $\eta_{\parallel}
\approx 3$ for either $\cos \theta_{\varphi} \approx 0$ or $c \gg
1$\,. Also, the second slow roll condition $|\eta_{\parallel}| \ll 1$
requires that $c < 1$ and $\cos \theta_{\varphi} < 0$\,. In fact,
$\eta_{\parallel} \approx 0$ is achieved for $c \approx - \cos
\theta_{\varphi} / 3$\,.

We next consider equation (\ref{eta_PP}) for $\eta_\perp$. Using the
entropic equation (\ref{Om_2}) and substituting $N$ from
(\ref{BasesRel}) gives:
\ben
\label{eta_perp_c}
\eta_{\perp} = \frac{\Omega}{H} = \frac{N_I V^I}{\dot{\sigma} H} = - \frac{(G^{KL}V_KV_L)^{\frac{1}{2}} \sin \theta_{\varphi}}{\dot{\sigma} H} = - \frac{\sin \theta_{\varphi}}{c} \,\,\, .
\een
In the second slow roll approximation we have \,$c \approx - \cos
\theta_{\varphi} / 3$\, (as explained above) and equation (\ref{eta_perp_c})
gives:
\ben
\label{eta_perp_tan}
\eta_{\perp}^2 \approx 9 \,\tan^2 \theta_{\varphi} \,\,\, .
\een
Hence the rapid turn condition $\eta_{\perp}^2 \gg 1$ is equivalent with
$\tan^2 \theta_{\varphi} \gg 1$ or, equivalently, with $\cos^2
\theta_{\varphi} \ll 1$\,. In view of (\ref{eta_par_c}), it follows that the
second slow roll condition $|\eta_{\parallel}| \ll 1$ requires $c^2
\ll 1$\,. Note that $c$ itself does not need to be very small. It
is enough to have $c \sim {\cal O} (10^{-1})$\,, in order to ensure
rapid turn and slow roll.

The discussion above shows that the slow roll and rapid turn regime, which is usually defined by:
\ben
\label{U_def_sr_rt}
\varepsilon \ll 1 \quad , \quad |\eta_{\parallel}| \ll 1 \quad , \quad \eta_{\perp}^2 \gg 1 \,\,\, , 
\een
is characterized equivalently by:
\ben
\label{SRRT}
\varepsilon \ll 1 \quad , \quad |\eta_{\parallel}| \ll 1 \quad , \quad c^2 \ll 1 \,\,\, . 
\een
We next derive the consistency condition on $V$ and $G_{IJ}$ in this
regime, supplemented by the additional requirements that:
\ben \label{xi_nu_cond}
|\xi| \ll 1 \quad {\rm and} \quad |\nu| \ll 1 \,\,\, ,
\een
where $\xi$ is the third order slow roll parameter in
(\ref{3rdOsr_par}), while $\nu$ is the relative rate of change of
$\eta_{\perp}$ as defined in (\ref{nu_def}). The purpose of the second
condition in (\ref{xi_nu_cond}) is to ensure that the inequality
(\ref{Large_eta_perp}) is satisfied for a prolonged period. (Note that $|\nu| \ll 1$ does not follow from the slow roll conditions in (\ref{U_def_sr_rt}).)

\subsection{The condition for sustained rapid turn with third order slow roll}

\noindent The equations of motion (\ref{EoM_sigma}) and (\ref{Om_2}) respectively imply (see \cite{HP,AGHPP,CCLBNZ}):
\beqan
\label{VTT_VTN}
\frac{V_{TT}}{3 H^2} &=& \frac{\Omega^2}{3 H^2} + \varepsilon + \eta_{\parallel} - \frac{\xi}{3} \,\,\, , \nn \\
\frac{V_{TN}}{H^2} &=& \frac{\Omega}{H} \left( 3 - \varepsilon - 2 \eta_{\parallel} + \nu \right) \,\,\, .
\eeqan
We will study the consequences of these relations for inflation with third order slow roll, defined by:
\ben
\label{SR3}
\varepsilon \,, \,|\eta_{\parallel}| \,, \,|\xi| \ll 1 \,\,\, ,
\een
while also imposing the condition:
\ben
\label{SCT}
|\nu| \ll 1 \,\,\, .
\een
The latter ensures a small rate of change of the large turn rate and hence longevity of the rapid turn regime. 

In the regime defined by (\ref{SR3})-(\ref{SCT}), equations (\ref{VTT_VTN})
imply:
\ben
\label{VTT_VTN_H}
V_{TT} \approx \frac{1}{9} \frac{V_{TN}^2}{H^2} \,\,\, .
\een
Recalling that $ 3 H^2 \approx V$ (see \eqref{H_V_sr}), relation (\ref{VTT_VTN_H}) can be written as:
\ben
\label{VTT_VTN_V}
V_{TT} \approx \frac{1}{3} \frac{V_{TN}^2}{V} \,\,\, .
\een
To extract conditions on $V$ and $G_{IJ}$\,, let us rewrite
(\ref{VTT_VTN_V}) in terms of $V_{nn}$, $V_{n \tau }$ and $V_{\tau
  \tau}$\,. Using (\ref{BasesRel}), we have:
\beqan
\label{VTN-Vntau}
\hspace*{-0.4cm}V_{TT} &=& T^I T^J \nabla_I V_J = V_{nn} \cos^2 \theta_{\varphi} + 2 V_{n \tau} \sin \theta_{\varphi} \cos \theta_{\varphi} + V_{\tau \tau} \sin^2 \theta_{\varphi} \,\,\, , \nn \\
\hspace*{-0.4cm}V_{TN} &=& T^I N^J \nabla_I V_J = \left( V_{\tau \tau} - V_{nn} \right) \cos \theta_{\varphi} \sin \theta_{\varphi} + V_{n \tau} \left( \cos^2 \theta_{\varphi} - \sin^2 \theta_{\varphi} \right) \,\, .
\eeqan
With the approximations (\ref{SRRT}), one has \,$\cos
\theta_{\varphi} \approx - 3 c$ \,due to (\ref{eta_par_c})\,. In that
case $\sin \theta_{\varphi} \approx s \sqrt{1 - 9 c^2}$\,, where $s =
\pm 1$\,. Hence (\ref{VTN-Vntau}) becomes:
\beqan
\label{VTN-Vntau_c}
\hspace*{-0.4cm}V_{TT} &\approx & 9 c^2 V_{nn} - 6 s c \,\sqrt{1-9c^2} \,V_{n \tau} + (1-9c^2) V_{\tau \tau} \,\,\, , \nn \\
\hspace*{-0.4cm}V_{TN} &\approx & - 3 s c \,\sqrt{1-9c^2} \left( V_{\tau \tau} - V_{nn} \right) - (1-18c^2) V_{n \tau} \,\,\, .
\eeqan
Since the different components of the Hessian of $V$ may be of
different orders in the small parameter $c$\,, we cannot conclude from
(\ref{VTN-Vntau_c}) that $V_{TT} \approx V_{\tau \tau}$ and $V_{TN}
\approx - V_{n \tau}$\,. Instead, we need two relations between $V_{nn}$, $V_{n
  \tau}$, $V_{\tau \tau}$ and $c$\,, in order to be able to solve for
$c$ and then extract a relation, that involves only the components of
the Hessian of $V$\,.

To achieve this, let us compute (\ref{VTT_VTN}) in terms of $c$ in the regime (\ref{SRRT}):
\beqan
\frac{V_{TT}}{3 H^2} &\approx & \frac{\Omega^2}{3 H^2} = \frac{\sin^2 \theta_{\varphi}}{3 \,c^2} \approx \frac{1}{3\,c^2} - 3 \,\,\, , \nn \\
\frac{V_{TN}}{H^2} &\approx & 3 \frac{\Omega}{H} = - \frac{3 \,\sin \theta_{\varphi}}{c} \approx - \frac{3 s}{c} \sqrt{1-9c^2} \,\,\, ,
\eeqan
where we used (\ref{eta_perp_c}) together with \,$\sin
\theta_{\varphi} \approx s \sqrt{1 - 9 c^2}$\,. Since $3 H^2 \approx V$ during slow roll, we have:
\ben
\label{VTN_c}
V_{TT} \approx \frac{V}{3\,c^2} - 3 V \qquad {\rm and} \qquad V_{TN} \approx - \frac{s}{c} \sqrt{1-9c^2}\,V \quad .
\een
These expressions satisfy (\ref{VTT_VTN_V}) automatically, as should be the case. 
Now substitute (\ref{VTN-Vntau_c}) in (\ref{VTN_c}). This gives the following two relations:
\beqan
\label{VTN-Vntau_V_c}
\frac{V}{3\,c^2} - 3 V &\approx & 9 c^2 V_{nn} - 6 s c \,\sqrt{1-9c^2} \,V_{n \tau} + (1-9c^2) V_{\tau \tau} \,\,\, , \nn \\
\frac{s}{c} \sqrt{1-9c^2}\,V &\approx & 3 s c \,\sqrt{1-9c^2} \left( V_{\tau \tau} - V_{nn} \right) + (1-18c^2) V_{n \tau} \,\,\, .
\eeqan
Our next goal is to eliminate the small parameter $c$\, without
making any assumptions about the order of any of the quantities $V$,
$V_{nn}$, $V_{n \tau}$, $V_{\tau \tau}$. We begin by solving for
$V_{n \tau}$ from the second equation in
(\ref{VTN-Vntau_V_c}):
\ben
\label{Vntau}
V_{n \tau} \,\approx \,\frac{s \sqrt{1-9c^2}}{(1-18c^2)} \left[ \frac{V}{c} - 3c \left( V_{\tau \tau} - V_{nn} \right) \right] \,\, .
\een
Substituting this in the first equation of (\ref{VTN-Vntau_V_c}) gives:
\ben
\label{FEq_interm}
(1-18c^2) \left( \frac{V}{3\,c^2} - 3 V \right) \approx - 6 V + 54 c^2 V - 9 c^2 V_{nn} + (1-9c^2) V_{\tau \tau} \,\,\, ,
\een
where we did {\em not} drop any term containing higher powers of $c$\,;
at intermediate steps there are terms with $c^4$\,, as well as other
$c^2$ terms, but they all cancel exactly. Since $(1-18c^2) \left(
\frac{V}{3\,c^2} - 3 V \right) = \frac{V}{3c^2} - 3V - 6V + 54 c^2
V$\, relation (\ref{FEq_interm}) becomes:
\be
\frac{V}{3\,c^2} - 3 V \,\approx \,- 9 c^2 V_{nn} + (1-9c^2) V_{\tau \tau} \,\,\, .
\ee
Thus:
\ben
\label{Vtautau}
V_{\tau \tau} \,\approx \,\frac{\frac{V}{3\,c^2} - 3 V + 9 c^2 V_{nn}}{1-9c^2} \,\,\, .
\een
Note that this expression for $V_{\tau \tau}$ was obtained without
neglecting any terms compared to (\ref{VTN-Vntau_V_c}). From (\ref{Vtautau}), we have:
\ben
\label{Vtt}
V_{\tau \tau} \approx \frac{V}{3c^2} + 9 c^2 V_{nn} + {\cal O} (c^4) \,\,\, .
\een 
Substituting (\ref{Vtautau}) in (\ref{Vntau}), we find:
\ben
\label{Vntau_2}
V_{n \tau} \approx \frac{3 s c V_{nn}}{\sqrt{1-9c^2}} \,\,\, ,
\een
where the terms with $V$ canceled exactly. We stress that
the result (\ref{Vntau_2}) was obtained without neglecting any further terms,
just like (\ref{Vtautau}). Now recall that in the rapid turn regime we have
$c^2 \ll 1$\,. Hence in this regime equations (\ref{Vtautau}) and (\ref{Vntau_2})
imply:
\ben
\label{Rels}
V_{\tau \tau} \approx \frac{V}{3 c^2} \qquad {\rm and} \qquad V_{n \tau} \approx 3sc V_{nn} \,\,\, .
\een
The second relation gives:
\ben
\label{c_VntauVnn}
c \approx \frac{s}{3} \frac{V_{n \tau}}{V_{nn}} \,\,\, .
\een
Substituting this into the first relation of (\ref{Rels}) gives:
\ben
\label{VCond}
3 V V_{nn}^2 \approx V_{n \tau}^2 V_{\tau \tau} \,\,\, .
\een
Since $V > 0$\,, the last relation implies, in particular, that we have: 
\ben \label{Vtt_pos}
V_{\tau \tau} > 0 \,\,\, .
\een

In view of (\ref{Vtt}), the Hessian of $V$ takes the following form to order $c^2$\,:
\be
\left[\begin{array}{cc} V_{\tau \tau} & ~~ V_{n \tau}\\ V_{n \tau} &
    ~~ V_{nn}\end{array}\right] \approx \left[\begin{array}{cc} 9 c^2
    V_{nn}+\frac{V}{3 c^2} & ~~~3scV_{nn}\\ 3scV_{nn} &
    ~~~V_{nn}\end{array}\right]~~.
\ee
The eigenvalues of the characteristic polynomial of this matrix are:
\be
\lambda_1 \approx V_{nn} \qquad {\rm and} \qquad \lambda_2 \approx \frac{V}{3 \,c^2} \quad ,
\ee
where we used $c^2 \ll 1$\,. Hence the Hessian of the potential has non-negative eigenvalues 
when:\footnote{The condition for non-tachyonic eigenvalues of the Hessian of $V$ was studied, 
for instance, in \cite{ACPRZ} in the context of supergravity. Note, however, that for 
rapid-turning models the conditions for stability of the trajectory are more subtle, 
and not necessarily incompatible with tachyonic eigenvalues; see \cite{CRS3}.}
\be
V_{nn} \ge 0 \,\,\, .
\ee
In view of (\ref{Rels}), the second eigenvalue is given by $\lambda_2 \approx
V_{\tau \tau}$\,. This is consistent with the fact that (\ref{c_VntauVnn}) implies (to leading order in the slow roll and rapid
turn regime) that the Hessian of $V$ is diagonal in the basis $(n,\tau)$\,.

From (\ref{VCond}) we have:
\ben
\label{Vtautau_1}
V_{\tau \tau} \approx 3 V \frac{V_{nn}^2}{V_{n \tau}^2} \,\,\, .
\een
Let us compare this with the expression for $V_{\tau \tau}$ in (\ref{Vtt}). Substituting (\ref{c_VntauVnn}) in (\ref{Vtt}) gives:
\ben
\label{Vtautau_2}
V_{\tau \tau} \approx \frac{V}{3\,c^2} + 9 c^2 V_{nn} \approx 3 V \frac{V_{nn}^2}{V_{n \tau}^2} + \frac{V_{n \tau}^2}{V_{nn}} \,\,\, .
\een
Using (\ref{Vtautau_1}) and (\ref{Vtautau_2}) we conclude that:
\ben \label{ConsCond_Ineq}
V_{\tau \tau} V_{nn} \gg V_{n \tau}^2 \,\,\, .
\een
Notice that the consistency condition (\ref{VCond}) depends not only on $V$
but also on the field-space metric $G_{IJ}$\,, as is clear from the
definitions of the basis vectors $n$ and $\tau$\,.

Finally, let us discuss the special case when $V_{n \tau} = 0$\,. In
this case, we must reconsider (\ref{VTN-Vntau_V_c}), since
(\ref{c_VntauVnn}) gives $c\approx 0$\,. So let us take $V_{n \tau}
= 0$ in (\ref{VTN-Vntau_V_c}):
\beqan
\label{Vttnn_V_c}
\frac{V}{3\,c^2} - 3 V &\approx & 9 c^2 V_{nn} + (1-9c^2) V_{\tau \tau} \,\,\, , \nn \\
\frac{s}{c} \sqrt{1-9c^2}\,V &\approx & 3 s c \,\sqrt{1-9c^2} \left( V_{\tau \tau} - V_{nn} \right) \,\,\, .
\eeqan
The second relation gives:
\be
V_{\tau \tau} \approx \frac{V}{3c^2} + V_{nn} \,\,\, .
\ee
Substituting this into the first equation in (\ref{Vttnn_V_c}) we obtain:
\ben \label{Vnn_Vnt0}
V_{nn} \approx 0 \,\,\, ,
\een
where we did not neglect any terms; everything except $V_{nn}$ canceled exactly. 
Thus, the consistency condition (\ref{VCond}) is satisfied again, although 
relation (\ref{ConsCond_Ineq}) is not valid anymore. 

To recapitulate the considerations of this Section: We have shown that (\ref{VCond}) 
is a necessary condition for the existence of a prolonged inflationary period in the 
slow-roll and rapid-turn regime defined by (\ref{U_def_sr_rt}) and (\ref{xi_nu_cond}). 
Note that, since $V_{nn} = n^I n^J \nabla_I V_J$ and similarly for $V_{n \tau}$ and 
$V_{\tau \tau}$\,, equation (\ref{VCond}) is a rather complicated PDE, whose coefficients 
depend on the scalar field-space metric $G_{IJ}$\,. Hence, solving it systematically, for 
instance in order to determine the potential for a given metric (or vice versa), is a very 
complicated problem. We certainly hope to report on progress in that regard, for some 
classes of scalar field metrics, in the future.

\section{The case of rotationally invariant metrics}
\setcounter{equation}{0}
\label{sec:rot}

In the literature on cosmological inflation the field-space metric
$G_{IJ}$ is often assumed to be rotationally invariant. Partly, this
is motivated by the desire for simplification. It is also the
case that many inflationary models which arise from string theory
compactifications have a scalar field metric of this type. In view of 
this, we will now specialize the consistency conditions derived in
the previous Section to the case of rotationally invariant metrics.

For such metrics, the consistency condition
extracted in the previous section can be made explicit upon using
local semigeodesic coordinates on the scalar manifold $(\cM,\cG)$, in
which the metric has the form:
\ben \label{Grotinv}
\dd s^2_G = \dd r^2 + h^2 (r) \dd\theta^2 \,\,\, 
\een
with $h (r) > 0$\,. In such coordinates, the non-vanishing Christoffel symbols are:
\ben
\Gamma^r_{\theta \theta} = - h h' \qquad , \qquad \Gamma^{\theta}_{r \theta} = \frac{h'}{h} \quad ,
\een
where the prime denotes derivation with respect to $r$.  Let us
compute the components of the Hessian, which enter the condition
(\ref{VCond}). Using (\ref{ndef}), (\ref{taudef}) and (\ref{Grotinv}),
we obtain:
\bea
V_{nn} &=& n^I n^J \nabla_I \nabla_J V \nn \\
&=& \frac{1}{(h^2 V_r^2 + V_{\theta}^2)} \left[ h^2 V_r^2 V_{rr} + 2 V_r V_{\theta} V_{r \theta} - \frac{h'}{h} V_r V^2_{\theta} + \frac{1}{h^2} V_{\theta}^2 V_{\theta \theta} \right] \,, \label{Vnn_rotinvG} \\
V_{n \tau} &=& n^I \tau^J \nabla_I \nabla_J V \nn \\
&=& \frac{h}{(h^2 V_r^2 + V_{\theta}^2)} \left[ V_r V_{\theta} \left( V_{rr} - \frac{1}{h^2} V_{\theta \theta} \right) + \left( \frac{1}{h^2} V^2_{\theta} - V_r^2 \right) V_{r \theta} - \frac{h'}{h^3} V^3_{\theta} \right] \,, \\
V_{\tau \tau} &=& \tau^I \tau^J \nabla_I \nabla_J V \nn \\
&=& \frac{h^4}{(h^2 V_r^2 + V_{\theta}^2)} \left[V_{\theta}^2 V_{rr} - 2 V_r V_{\theta} \left( V_{r \theta} - \frac{h'}{h} V_{\theta} \right) + V_r^2 (V_{\theta \theta} + h h' V_r) \right] \,. \label{Vtt_rotinvG}
\eea
The result from substituting (\ref{Vnn_rotinvG})-(\ref{Vtt_rotinvG})
in (\ref{VCond}) is too messy to be illuminating without further
specialization. However, the very existence of this highly nontrivial
relation implies that it is not easy to achieve a long lasting
period of rapid turn with slow roll. In other words, one can expect that obtaining at
least 50-60 e-folds of slow-roll rapid-turn inflation is possible 
  only for special choices of scalar potentials and field-space
metrics.

Expressions (\ref{Vnn_rotinvG})-(\ref{Vtt_rotinvG}) simplify
enormously for $\theta$-independent scalar potentials $V$. In that case,
one finds:
\be
V_{nn} = V_{rr} \quad , \quad V_{n \tau} = 0 \quad , \quad V_{\tau \tau} = h^3 h' V_r \quad .
\ee
If we assume that $V_{rr} \neq 0$\, then the consistency condition
(\ref{VCond}) cannot be satisfied anywhere in field space, regardless
of the specifics of the model. This agrees with the
conclusion, reached in (\ref{Vnn_Vnt0}) for the case with $V_{n \tau}
= 0$\,, that $V_{nn}$ has to vanish in that case for consistency.

The discussion in this section shows in particular that sustained
rapid turn with third order slow roll cannot be realized in models
with a rotationally invariant scalar potential, regardless of the form
of the field-space trajectory.\footnote{Note that in the solutions of
  \cite{ADGW} (which arise from a potential of the form $V = V (r)$) one
  has $\eta_{\parallel}^{DE} \sim {\cal O} (1)$\,. Of course, there
  are no observational constraints on the parameter $\eta_{\parallel}$
  for dark energy. Also, as discussed in \cite{ADGW}, the mass of the
  corresponding entropic perturbation is always positive in those
  solutions, so there is no tachyonic instability at the perturbative
  level. However, they do {\em not} provide counterexamples to the
  conclusion of this section that long-term slow-roll rapid-turn
  inflation cannot occur with a rotationally invariant potential.}

\subsection{Slow roll consistency condition for circular trajectories} \label{SR_CC_CT}

\noindent An Ansatz that is often used in the literature to look for
sustained slow-roll and rapid-turn solutions which might lead to the
phenomenologically desirable 50-60 or so e-folds of inflation
consists of taking $V_{\theta} \neq 0$ and considering near-circular
field-space trajectories. Thus one assumes $r \approx \const$\,, which
more precisely means that one approximates $\dot{r} \approx 0$ and
$\ddot{r} \approx 0$\,. This approximation was used for example in
\cite{ACPRZ} as well as in many other references on rapid turn
inflation.

However, there is a consistency condition for compatibility
of this approximation with the equations of motion. To see this, let
us first specialize the form of the field equations to the case of a
rotationally invariant field space metric. For $G_{IJ}$ given by
(\ref{Grotinv}), the equations for the scalar fields (\ref{EoM_sc})
become:
\bea
\ddot{r} - h h' \dot{\theta}^2 + 3 H \dot{r} + V_r &=& 0 \,\,\, , \label{EoMr} \\
\ddot{\theta} + 2 \frac{h'}{h} \dot{r} \dot{\theta} + 3 H \dot{\theta} + \frac{1}{h^2} V_{\theta} &=& 0 \label{EoMth} \,\,\, .
\eea

We next approximate $\dot{r} \approx 0$ and $\ddot{r} \approx
0$\,. In this case, the second slow roll parameter takes
the form:
\ben \label{eta_par_rth}
\eta_{\parallel} = - \frac{ \dot{r} \ddot{r} + h h' \dot{r} \dot{\theta}^2 + h^2 \dot{\theta} \ddot{\theta}}{ H ( \dot{r}^2 + h^2 \dot{\theta}^2 )} \approx - \frac{\ddot{\theta}}{H \dot{\theta}} \,\,\, .
\een
Imposing the slow roll conditions $\varepsilon \ll 1$ and
$|\eta_{\parallel}| \ll 1$\, within the approximations 
$\dot{r}\approx 0$ and $\ddot{r} \approx 0$
for a sustained near-circular trajectory, we find that
(\ref{EoMr})-(\ref{EoMth}) reduce to:\footnote{Note that, without the
assumption $r \approx \const$, the second slow roll condition
$|\eta_{\parallel}| \ll 1$ {\it does not} amount to neglecting
$\ddot{\theta}$ in the equations of motion, as is clear from
(\ref{eta_par_rth}).}
\bea
-h h' \dot{\theta}^2 + V_r &\approx& 0 \,\,\, \label{EoMr_ca} \\
3 H \dot{\theta} + \frac{V_{\theta}}{h^2} &\approx& 0 \label{EoMth_ca} \,\,\, .
\eea
From the second equation in (\ref{EinstEqs}) we also have:
\ben \label{H_EE_sr}
3H^2 \approx V \,\,\, ,
\een
because $\varepsilon \ll 1$ in the slow roll regime. Notice that
(\ref{EoMth_ca}) does not admit solutions with $\dot{\theta} \neq 0$
for rotationally invariant potentials. Equation (\ref{EoMr_ca}) gives:
\ben \label{th_d_sq_ca}
\dot{\theta}^2 \approx \frac{V_r}{hh'} \,\,\, ,
\een
whereas (\ref{EoMth_ca}) implies:
\ben \label{H_ca}
H \approx - \frac{V_{\theta}}{3h^2 \dot{\theta}} \,\,\, .
\een
Substituting (\ref{H_ca}) in (\ref{H_EE_sr}) gives:
\ben \label{V_Vth_ca}
\frac{V_{\theta}^2}{3h^4\dot{\theta}^2} \approx V \,\,\, .
\een
Finally, substituting (\ref{th_d_sq_ca}) in (\ref{V_Vth_ca}), 
we find (in agreement with \cite{CRS2}):
\ben \label{h-V_rel}
\frac{h'}{3h^3} \frac{V_{\theta}^2}{V_r} \approx V \,\,\, .
\een
This relation need not be satisfied everywhere in field space. But it
has to be (approximately) satisfied along a
slowly rolling inflationary solution of the background equations of
motion with a near circular field space trajectory.

To illustrate the usefulness of the consistency condition
(\ref{h-V_rel}), let us apply it to an example that was considered in
\cite{ACPRZ} when looking for rapid turn solutions in the slow roll
regime. To do that, we first note that the field-space metric of
\cite{ACPRZ} has the form:
\ben
\label{G_rh_th}
ds^2 = h^2(\rho) \left( d\rho^2 + d\theta^2\right) \,\,\, .
\een
This is related to (\ref{Grotinv}) via the field redefinition:
\be
\frac{d r}{d \rho} = h \,\,\, .
\ee
Relation (\ref{h-V_rel}) has the same form in terms of the variable $\rho$, namely:
\ben
\label{h_hr-V_Vd}
\frac{h_{\rho}}{3 h^3} \frac{V_{\theta}^2}{V_{\rho}} = V \,\,\, ,
\een
where $h_{\rho} \eqdef \pd_{\rho} h$ and $V_{\rho} \eqdef \pd_{\rho} V$\,.

Now we turn to investigating whether (\ref{h_hr-V_Vd}) can be satisfied in
the no-scale inspired rapid-turn model considered in \cite{ACPRZ}. In that case,
the functions $h$ and $V$ are:
\ben
\label{hV_ns}
h_{ns} = \frac{3 \alpha}{2 \rho^2} \qquad , \qquad V_{ns} = \frac{p_1^2 \theta^2 + (p_0+p_1 \rho )^2}{8^{\alpha} \rho^{3 \alpha}} \quad ,
\een
where $\alpha,p_0,p_1 = \const$\,. Rapid turn is achieved 
by taking the parameter $\alpha$ to be small. In that
limit, the leading term in the potential is:
\ben
\label{RHS}
V_{ns} = p_1^2 \theta^2 + (p_0+p_1 \rho )^2 \,\,\, .
\een
On the other hand, the leading term in the expression on the left-hand side of (\ref{h_hr-V_Vd}) is:
\ben
\label{LHS}
\frac{h_{\rho}}{3 h^3} \frac{V_{\theta}^2}{V_{\rho}}\bigg|_{ns} = - \frac{16}{27} \frac{p_1^3 \rho^3}{(p_0+p_1 \rho )} \frac{\theta^2}{\alpha^2} \,\,\, .
\een
Now recall that $\rho \approx \const$ in the present context.
Therefore, for finite $\theta$ and small $\alpha$, the
expression (\ref{LHS}) diverges while (\ref{RHS}) does not. Hence the
two expressions could be (almost) equal numerically only for $\theta
\ll 1$ and such that $\theta \sim {\cal O} (\alpha)$\,. However, these
conditions are not compatible with the slow roll conditions, which for 
$\dot{\rho} \approx 0$ are equivalent with demanding that $\epsilon_T \ll 1$ 
and $\eta_T \ll 1$\, (see \cite{ACPRZ}).

To show this, let us consider in more detail the slow roll parameters
$\epsilon_T$ and $\eta_T$\,, which are given by:
\ben \label{EpT_EtT_ca}
\epsilon_T \approx \frac{1}{2 h^2} \left( \frac{V_{\theta}}{V} \right)^2 \quad \, {\rm and} \qquad \eta_T \approx \frac{1}{h^2} \frac{V_{\theta \theta}}{V}
\een
in the approximation $\rho \approx \const$ \,(see \cite{ACPRZ}). For the
functions in (\ref{hV_ns}), we find:
\ben
\epsilon_T|_{ns} \approx \frac{8}{9} \, \frac{p_1^4 \rho^4}{\left[p_1^2\theta^2 + (p_0+p_1 \rho )^2\right]^2} \, \frac{\theta^2}{\alpha^2}
\een
and:
\ben \label{eta_T_ca}
\eta_T|_{ns} \approx \frac{8}{9} \, \frac{p_1^2 \rho^4}{\left[ p_1^2\theta^2 + (p_0+p_1 \rho )^2 \right]} \, \frac{1}{\alpha^2} \quad ,
\een
where we did not neglect any subleading terms in small $\alpha$. These
equalities are approximate only due to the approximations in
(\ref{EpT_EtT_ca}). Since $p_0,p_1$ and $\rho$ are finite constants,
the slow roll condition $\epsilon_T \ll 1$ can only be satisfied if
$\theta \ll \alpha$\,. However, this contradicts the condition $\theta
\sim {\cal O} (\alpha)$ found above. Conversely, requiring $\epsilon_T
\ll 1$ necessarily violates the consistency condition
(\ref{h_hr-V_Vd}), which encodes compatibility with the equations of
motion. Therefore, the above method of achieving slow roll and rapid
turn is incompatible with the equations of motion. We suspect that the
putative inflationary trajectories found in loc. cit solve only the
scalar field equations, but not the Einstein equations. The situation
is even worse for the slow roll condition $\eta_T \ll 1$\,, since for
small $\alpha$ (i.e. for rapid turn) the expression in
(\ref{eta_T_ca}) diverges regardless of any comparison between the
magnitudes of $\alpha$ and $\theta$. In other words, slow roll and
rapid turn are incompatible with each other in this context.

In the example of the EGNO model considered in \cite{ACPRZ}, the
situation is conceptually the same as in the no-scale inspired model
discussed above, although the relevant expressions are more
cumbersome.

The lesson from this section is that it is not reliable to look for
rapid turn solutions by tuning parameters in known slow-roll slow-turn
solutions. The conceptual reason is that the latter are approximate
solutions of the equations of motion {\it only} in some parts of their
parameters spaces, and not for arbitrary parameter values (because
they satisfy the equations of motion up to error terms whose magnitude
depends on the parameters). As a consequence, arbitrary variations of the
parameters in such solutions can violate the approximations within which they
solve (approximately) the equations of motion.\footnote{For further 
illustration of the usefulness of the consistency conditions studied here 
(and, in particular, of the constructive role they can play in model building), 
see the discussion in Appendix \ref{QSFI}.}

\section{A unifying approximation for rapid turn models}
\setcounter{equation}{0}
\label{sec:Bjorkmo}

A number of prominent rapid turn models of inflation (including
\cite{AB,SM,GSRPR,CRS,ACIPWW}) can be viewed as special cases of the
unifying framework proposed in \cite{TB}. The latter uses the adapted frame
and does not rely on specific choices of
potential $V$ and metric $G_{IJ}$\,. It thus appears to provide a wide
class of realizations of rapid turn inflation. It turns out
that the considerations of \cite{TB} rely on an approximation that is
less general than slow roll, as will become evident below. We will
also show that imposing this approximation leads to
rather nontrivial consistency conditions for compatibility
with the equations of motion.

\subsection{Equations of motion in adapted frame}

\noindent Let us begin by rewriting the equations of motion
(\ref{EoM_sc}) in the basis $(n,\tau)$\,. For this, we first
expand $\dot{\varphi}^I = \dot{\sigma} T^I$ as:
\ben \label{phiI_ntau}
\dot{\varphi}^I = v_n n^I + v_{\tau} \tau^I \,\,\, ,
\een
where (see (\ref{BasesRel})):
\ben
\label{dtphi_ntau}
v_n \eqdef n_I \dot{\varphi}^I = \dot{\sigma} \cos \theta_{\varphi} \qquad {\rm and} \qquad v_{\tau} \eqdef \tau_I \dot{\varphi}^I = \dot{\sigma} \sin \theta_{\varphi} \,\,\, .
\een
Therefore:
\ben
\label{DtphiI}
D_t \dot{\varphi}^I \eqdef \dot{\varphi}^J \nabla_J \dot{\varphi}^I = v_n \nabla_n \dot{\varphi}^I + v_{\tau} \nabla_{\tau} \dot{\varphi}^I \,\,\, .
\een
To compute the right hand side, note that (see Appendix \ref{DAdFr}):
\beqan
\label{nabla_ntau}
\nabla_n n = \mu \tau &,& \nabla_n \tau = - \mu n \,\,\, , \nn \\
\nabla_{\tau} \tau = \lambda n &,& \nabla_{\tau} n = - \lambda \tau \,\,\, ,
\eeqan
where:
\ben
\label{la_mu}
\lambda = - \frac{V_{\tau \tau}}{\sqrt{G^{IJ} V_I V_J}} \quad , \quad \mu = \frac{V_{n \tau}}{\sqrt{G^{IJ} V_I V_J}} \,\,\, .
\een
Substituting (\ref{phiI_ntau}) and (\ref{nabla_ntau}) in (\ref{DtphiI}) gives:
\ben
\label{Dtphi_ntau}
D_t \dot{\varphi}^I = \left( \dot{v}_n - \lambda v_{\tau}^2 - \mu v_n v_{\tau} \right) n^I +
\left( \dot{v}_{\tau} + \mu v_n^2 + \lambda v_n v_{\tau} \right) \tau^I \,\,\, ,
\een
where we used the relation $\dot{\varphi}^I \nabla_I v_n =
\dot{\varphi}^I \pd_I v_n = \dot{v}_n$\,.  Also notice
that (\ref{ndef}) implies:
\ben
\label{Vn}
G^{IJ} V_J \,= \,\sqrt{G^{KL}V_K V_L} \,\,\, n^I \,\,\, .
\een
Using (\ref{phiI_ntau}), (\ref{Dtphi_ntau}) and (\ref{Vn}) we find
that the projections of (\ref{EoM_sc}) along $n$ and $\tau$
give the following equations respectively:
\beqan
\label{EoMs_ntau}
\dot{v}_n - \lambda v_{\tau}^2 - \mu v_n v_{\tau} + 3H v_n + \sqrt{V_I V^I} &=& 0 \,\,\, , \nn \\
\dot{v}_{\tau} + \mu v_n^2 + \lambda v_n v_{\tau} + 3H v_{\tau} &=& 0 \,\,\, .
\eeqan
Now let us consider the approximation within which these equations of motion were studied in \cite{TB}.

\subsection{Adapted frame parameters}

\noindent To discuss the approximation of \cite{TB,BM}, we introduce the following parameters:
\ben
\label{eta_n_eta_t}
f_n \eqdef - \frac{\dot{v}_n}{H v_n} \qquad , \qquad f_\tau \eqdef - \frac{\dot{v}_{\tau}}{H v_{\tau}} \quad .
\een
These quantities are adapted frame analogues of the slow roll
parameter $\eta_{\parallel}$\,. However, they do {\em not} play the same
role. To see this, let us write $\eta_{\parallel}$ in terms of $f_n$
and $f_{\tau}$\,. For this, notice first that:
\ben
\label{sigma_phi_n_phi_t}
\dot{\sigma}^2 = G_{IJ} \dot{\varphi}^I \dot{\varphi}^J = v_n^2 + v_{\tau}^2 \,\,\, ,
\een
where we used (\ref{phiI_ntau}). Using the first relation in
(\ref{eta_PP}) together with (\ref{dtphi_ntau}) and
(\ref{sigma_phi_n_phi_t}), we find:
\ben
\label{eta_par_n_t}
\eta_{\parallel} \,= \,\cos^2 \theta_{\varphi} \,f_n + \sin^2 \theta_{\varphi} \,f_\tau \,\,\, .
\een
The approximation regime in \cite{TB} is obtained by imposing the conditions:
\ben
\label{eta_n_eta_t_sr}
|f_n | \ll 1 \qquad , \qquad |f_\tau| \ll 1 \quad ,
\een
or more precisely \,$\dot{v}_{n,\tau} \sim {\cal O} (\varepsilon) H
v_{n,\tau}$ .\footnote{Note that this assumption is written in \cite{TB} as $\ddot{\varphi}_{n,\tau} \sim {\cal O} (\varepsilon) H \dot{\varphi}_{n,\tau}$\,, where the components $\dot{\varphi}_{n,\tau}$ are defined via $\dot{\varphi}^I = \dot{\varphi}_n n^I + \dot{\varphi}_{\tau} \tau^I$ and the quantity  $\ddot{\varphi}_{n,\tau}$ is the time-derivative of $\dot{\varphi}_{n,\tau}$\,. However, this notation is misleading as $\dot{\varphi}_{n,\tau}$ itself is not a time derivative (but a projection of $\dot{\varphi}^I$), and consequently $\ddot{\varphi}_{n,\tau}$ is not a second time derivative either. To avoid any resulting confusion, we are intentionally using different notation compared to \cite{TB}.} However, relation (\ref{eta_par_n_t}) shows that,
although the inequalities (\ref{eta_n_eta_t_sr}) imply the second slow
roll condition $|\eta_{\parallel}| \ll 1$\,, the latter does {\em not}
imply (\ref{eta_n_eta_t_sr}). Hence the considerations of \cite{TB}
encompass only a particular subset of trajectories which satisfy the
second slow roll condition. Indeed, a more general possibility is the
following. Since rapid turn requires $\cos^2 \theta_{\varphi} \ll
1$\, (as discussed below relation \eqref{eta_perp_tan}), one could satisfy the
slow roll condition $|\eta_{\parallel}| \ll 1$ by having $|f_n| \sim
{\cal O} (1)$ and $|f_\tau| \ll 1$\,. It would be very interesting to
investigate in more detail this new regime. In particular, its
existence seems to suggest that, counter-intuitively, one can have
slow roll while moving relatively fast along the gradient of the
potential.

Let us now compute the quantities $f_n$ and $f_\tau$ on solutions of
the equations of motion. Substituting $\dot{v}_n$ and $\dot{v}_{\tau}$
from (\ref{EoMs_ntau}) in (\ref{eta_n_eta_t}) gives:
\beqa
f_n &=& 3 + \frac{\sqrt{V_I V^I} - \lambda v^2_{\tau} - \mu v_n v_{\tau} }{H v_n} \quad , \nn \\
f_\tau &=& 3 + \frac{\mu v_n^2 + \lambda v_n v_{\tau} }{H v_{\tau}} \quad .
\eeqa
Using (\ref{c_def}), (\ref{dtphi_ntau}) and (\ref{la_mu}), the last expressions become:
\beqan
\label{eta_n_eta_t_c_trig}
f_n &=& 3 + \frac{1}{c \cos \theta_{\varphi}} - \frac{c}{H^2} \tan \theta_{\varphi} (- V_{\tau \tau} \sin \theta_{\varphi} + V_{n \tau} \cos{\theta}_{\varphi} ) \quad , \nn \\
f_\tau &=& 3 + \frac{c}{H^2 \tan \theta_{\varphi} } (- V_{\tau \tau} \sin \theta_{\varphi} + V_{n \tau} \cos{\theta}_{\varphi} ) \quad .
\eeqan
Notice that (\ref{eta_par_c}) and (\ref{eta_perp_c}) can be written respectively as:
\ben
\label{cos_sin_c}
\cos \theta_{\varphi} = c (\eta_{\parallel} - 3) \quad , \quad \sin \theta_{\theta} = - c \eta_{\perp} \quad ,
\een
which imply:
\beqan
\label{c_eta_perp_par}
c^2 = \frac{1}{\eta_{\perp}^2 + (\eta_{\parallel}-3)^2} \quad .
\eeqan
Using (\ref{cos_sin_c})-(\ref{c_eta_perp_par}) in (\ref{eta_n_eta_t_c_trig}), we find:
\beqa
f_n &=& \eta_{\parallel} + \frac{\eta_{\perp}^2}{\eta_{\parallel} - 3} + \frac{\eta_{\perp} \left[ V_{\tau \tau} \eta_{\perp} + V_{n \tau} (\eta_{\parallel}-3) \right] }{H^2 \,(\eta_{\parallel} - 3) \left[ \eta_{\perp}^2 + (\eta_{\parallel} - 3)^2 \right]} \quad , \nn \\
f_\tau &=& 3 - \frac{(\eta_{\parallel} - 3) \left[ V_{\tau \tau} \eta_{\perp} + V_{n \tau} (\eta_{\parallel} - 3) \right]}{H^2 \,\eta_{\perp} \left[ \eta_{\perp}^2 + (\eta_{\parallel} - 3)^2 \right]} \quad .
\eeqa
Imposing the first and second slow roll conditions, these expressions simplify to:
\beqan
\label{eta_perp_par_SR}
f_n &\approx& - \frac{\eta_{\perp}^2}{3} - \frac{\eta_{\perp} \left( V_{\tau \tau} \eta_{\perp} - 3 V_{n \tau} \right)}{V \left( \eta_{\perp}^2 + 9 \right)} \quad , \nn \\
f_\tau &\approx& 3 + \frac{9 \left( V_{\tau \tau} \eta_{\perp} - 3 V_{n \tau} \right)}{V \,\eta_{\perp} \left( \eta_{\perp}^2 + 9 \right)} \quad .
\eeqan
Notice that relations (\ref{eta_perp_par_SR}) imply:
\ben
\label{rel_eta_n_t}
f_\tau \approx - \frac{9}{\eta_{\perp}^2} \,f_n \,\,\, .
\een
This is in line with our observation above that one can have slow roll
and rapid turn with $|f_n| \sim {\cal O} (1)$ and $|f_\tau|
\ll 1$\,.

\subsection{Consistency conditions}

\noindent We next derive consistency conditions for the approximation given by (\ref{eta_n_eta_t_sr}). 
In view of (\ref{rel_eta_n_t}), that approximation holds when $|f_n| \ll 1$\,. 
In this case, (\ref{eta_perp_par_SR}) implies:
\ben
\label{cubic_Eq}
V \eta_{\perp} (\eta_{\perp}^2 + 9) \approx 3 (3 V_{n \tau} - V_{\tau \tau} \eta_{\perp}) \,\,\, .
\een
We will show below that combining this approximate equality with the
slow roll condition $\varepsilon \ll 1$ leads to a constraint on the
potential. 

Recall from (\ref{ep_kappa}) that $\varepsilon \ll 1$ is
equivalent with $\kappa \ll 1$\,. Using (\ref{c_def}) and (\ref{c_eta_perp_par}), we compute:
\be
\dot{\sigma}^2 = \frac{V_I V^I}{H^2 \left[ \eta_{\perp}^2 + (\eta_{\parallel} - 3)^2 \right]} \,\,\, .
\ee
Substituting this in (\ref{kappa_def}), we find that the first slow roll condition $\kappa \ll 1$ takes the form:
\be
\frac{V_I V^I}{H^2 \left[ \eta_{\perp}^2 + (\eta_{\parallel} - 3)^2 \right] 2 V} \ll 1 \,\,\, .
\ee
In the slow roll regime this reduces to:
\ben
\label{Eta_perp_ineq}
\eta_{\perp}^2 + 9 \gg \frac{3}{2} \frac{V_I V^I}{V^2} \,\,\, .
\een
Requiring compatibility between (\ref{Eta_perp_ineq}) and
(\ref{cubic_Eq}) imposes a constraint on the potential. Namely, there
must be at least one solution of the depressed cubic equation
(\ref{cubic_Eq}) for $\eta_{\perp}$\, which satisfies the inequality
(\ref{Eta_perp_ineq}).

Let us discuss the resulting constraint in more detail. For
convenience, we rewrite (\ref{cubic_Eq}) as:
\ben
\label{Cubic_pq}
\eta_{\perp}^3 + p \eta_{\perp} + q \approx 0 \,\,\, ,
\een
where:
\be
p \eqdef 9 + 3 \frac{V_{\tau \tau}}{V} \quad , \quad q \eqdef - 9 \frac{V_{n \tau}}{V} \,\,\, .
\ee
The discriminant of this equation is:
\be
\Delta = - 4 p^3 - 27 q^2 \,\,\, .
\ee
If $\Delta \ge 0$\,, then (\ref{Cubic_pq}) has three real roots. To
ensure that at least one of them satisfies (\ref{Eta_perp_ineq}), it
is enough to require the largest root to satisfy it.
According to Viete's formula, the three roots of the depressed cubic
can be written as:
\be
\eta_{\perp} = A \cos \left( \frac{1}{3} \arccos B + \frac{2 \pi}{3} k \right) \quad , \quad k = 0,1,2 \,\,\, ,
\ee
where $A \eqdef \sqrt{\frac{4 |p|}{3}}$ \,and \,$B
\eqdef \sqrt{\frac{27 q^2}{4 |p|^3}}$\,. Clearly, the largest root
belongs to the interval $[\frac{A}{2},A]$\,. Hence requiring that it
satisfy (\ref{Eta_perp_ineq}) amounts to the condition $9 +
\frac{A^2}{4} \gg \frac{3}{2} \frac{V_I V^I}{V^2}$\,, namely:
\ben \label{CCondDp}
12 + \frac{V_{\tau \tau}}{V} \gg \frac{3}{2} \frac{V_I V^I}{V^2} \,\,\, .
\een
If $\Delta < 0$\,, then (\ref{Cubic_pq}) has a single real root, which is given by Cardano's formula:
\be
\eta_{\perp *} = \sqrt[3]{-\frac{q}{2}+\sqrt{\frac{q^2}{4}+\frac{p^3}{27}}}+\sqrt[3]{-\frac{q}{2}-\sqrt{\frac{q^2}{4}+\frac{p^3}{27}}} \,\,\, .
\ee
Hence (\ref{Eta_perp_ineq}) becomes:
\ben \label{CCondDn}
\eta_{\perp *}^{2} + 9 \gg \frac{3}{2} \frac{V_I V^I}{V^2} \,\,\, .
\een

Formulas (\ref{CCondDp}) and (\ref{CCondDn}) are complicated relations
between $V$ and $G_{IJ}$\,. For any given field-space metric
$G_{IJ}$\,, this means that there is a highly nontrivial constraint on
the potential which limits significantly the choices of $V$ that are
compatible with solutions of the equations of motion in the slow roll
and rapid turn regime.

\section{The characteristic angle}
\setcounter{equation}{0}
\label{sec:charangle}

In this section we investigate the behavior of the characteristic
angle $\theta_{\varphi}$\,, which relates the oriented frames $(T,N)$
and $(n,\tau)$ according to equation (\ref{BasesRel}). For that
purpose, we will rewrite the adapted-frame equations of motion
(\ref{EoMs_ntau}) in a manner that leads to an equation determining
$\theta_{\varphi} (t)$\,. Although that equation cannot be solved
exactly without specifying the form of $V$ and $G_{IJ}$\,, it
nevertheless leads to a valuable universal insight about the duration
of the rapid turn regime.

To obtain an equation for $\theta_{\varphi} (t)$\,, let us substitute
(\ref{dtphi_ntau}) in the second equation of (\ref{EoMs_ntau}). The
result is:
\ben
\label{th_d}
\dot{\theta}_{\varphi} \cos \theta_{\varphi} + (3 - \eta_{\parallel}) H \sin \theta_{\varphi} + \dot{\sigma} (\mu  \cos^2 \theta_{\varphi} + \lambda \sin \theta_{\varphi} \cos \theta_{\varphi} ) = 0 \,\,\, ,
\een
where we used that \,$\ddot{\sigma} = - H \dot{\sigma}
\eta_{\parallel}$ \,in accordance with (\ref{eta_PP}). It is convenient to
rewrite (\ref{th_d}) as:
\ben
\label{th_d_E1}
\dot{\theta}_{\varphi} + (3 - \eta_{\parallel}) H \tan \theta_{\varphi} + \dot{\sigma} (\mu  \cos \theta_{\varphi} + \lambda \sin \theta_{\varphi} ) = 0 \,\,\, .
\een
Similarly, the first equation in (\ref{EoMs_ntau}) takes the form:
\ben
\label{th_d_E2}
\dot{\theta}_{\varphi} - (3 - \eta_{\parallel}) H \cot \theta_{\varphi} + \dot{\sigma} (\mu  \cos \theta_{\varphi} + \lambda \sin \theta_{\varphi} ) - \frac{H}{c \,\sin \theta_{\varphi}} = 0 \,\,\, ,
\een
where we used the relation $\sqrt{V_I V^I} = \frac{\dot{\sigma} H}{c}$\,,
in accordance with (\ref{c_def}). In the slow roll regime one has
\,$|\eta_{\parallel}| \!\ll \!1$ \,and \,$c \!\approx \!\!- \cos
\theta_{\varphi} / 3$ \,. Hence in the slow roll approximation
equation (\ref{th_d_E2}) coincides with (\ref{th_d_E1}). In
general, the difference between equations (\ref{th_d_E2}) and
(\ref{th_d_E1}) is precisely the on-shell relation (\ref{eta_par_c})
between $\eta_{\parallel}$, $\theta_{\varphi}$ and $c$\,.

Let us now study in more detail equation (\ref{th_d_E1}) in the slow
roll approximation. Using (\ref{c_def}) and $c
\approx - \cos \theta_{\varphi} / 3$\,, we have:
\be
\dot{\sigma} \approx - \frac{\cos \theta_{\varphi}}{3} \frac{\sqrt{V_I V^I}}{H} \,\,\, .
\ee
Substituting this together with (\ref{H_V_sr}) and (\ref{la_mu}) in
(\ref{th_d_E1}), we find that during slow roll one has:
\ben \label{Char_ang_Vnt_Vtt}
\dot{\theta}_{\varphi} + \sqrt{3 V} \tan \theta_{\varphi} - \frac{1}{\sqrt{3V}} \left( V_{n \tau} \cos^2 \theta_{\varphi} - V_{\tau \tau} \cos \theta_{\varphi} \sin \theta_{\varphi} \right) \approx 0 \,\,\, .
\een
Recall that rapid turn requires $\cos^2 \theta_{\varphi} \ll
1$\,, although $\cos \theta_{\varphi}$ itself need not be very
small. So assuming that $V_{n \tau}$ is of order $V_{\tau \tau}$ or
smaller, which is reasonable in view of (\ref{ConsCond_Ineq}), we can
simplify (\ref{Char_ang_Vnt_Vtt}) to:
\ben \label{Char_ang_Vtt}
\dot{\theta}_{\varphi} + \sqrt{3 V} \tan \theta_{\varphi} + \frac{1}{\sqrt{3V}} V_{\tau \tau} \cos \theta_{\varphi} \sin \theta_{\varphi} \approx 0 \,\,\, .
\een
When the $V_{\tau \tau}$ term is negligible, the last equation can be written in the nice form:
\ben \label{Th_H_c}
\dot{\theta}_{\varphi} \approx \pm \frac{H}{c} \,\,\, ,
\een
where we used the fact that \,$\cos \theta_{\varphi} \approx - 3c$ \,and
\,$\sin \theta_{\theta} \approx \pm 1$ \,during slow roll; of course,
one should keep in mind that the parameter $c$\, defined in
(\ref{c_def}) is not a constant along the field-space trajectory of an
inflationary solution.

Now let us consider one by one the cases when the $V_{\tau \tau}$
term in (\ref{Char_ang_Vtt}) is negligible and when it is not. In the
first case, we have:
\ben \label{Th_vph_lead}
\dot{\theta}_{\varphi} + \sqrt{3 V} \tan \theta_{\varphi} \,\approx \,0 \,\,\, .
\een
To gain insight in the behavior of $\theta_{\varphi} (t)$, let us
assume that, at least initially, the inflationary trajectory is close
to a level set of the potential\footnote{In view of (\ref{BasesRel}),
this is equivalent with $|\theta_{\varphi} (0)| \approx
\frac{\pi}{2}$\,.}, in other words that $V \approx \const$ along (an
initial part of) the trajectory. Then the solution of
(\ref{Th_vph_lead}) is given by:
\ben \label{th_varph_cr_b}
\sin \theta_{\varphi} \,\approx \,\pm \,\tilde{C} \,e^{- \,\sqrt{3V} \,t} \,\,\, ,
\een 
where $\tilde{C} = \const >0$\,. This means that $|\sin
\theta_{\varphi}|$ tends to zero with $t$\,, with a characteristic
timescale determined by $T = \frac{1}{\sqrt{3V}}$\,. This
is of course a rather crude estimate, which relies on neglecting the $V_{\tau
\tau}$ term and on the assumption $V \approx \const$\,. Nevertheless,
relation (\ref{th_varph_cr_b}) gives useful intuition. Note that, if the
initial part of the trajectory were not along a level set of the
potential, then $|\sin \theta_{\varphi}|$ would decay even faster.

\begin{figure}[t]
\centering
\begin{minipage}{.49\textwidth}
\centering  \includegraphics[width=.89\linewidth]{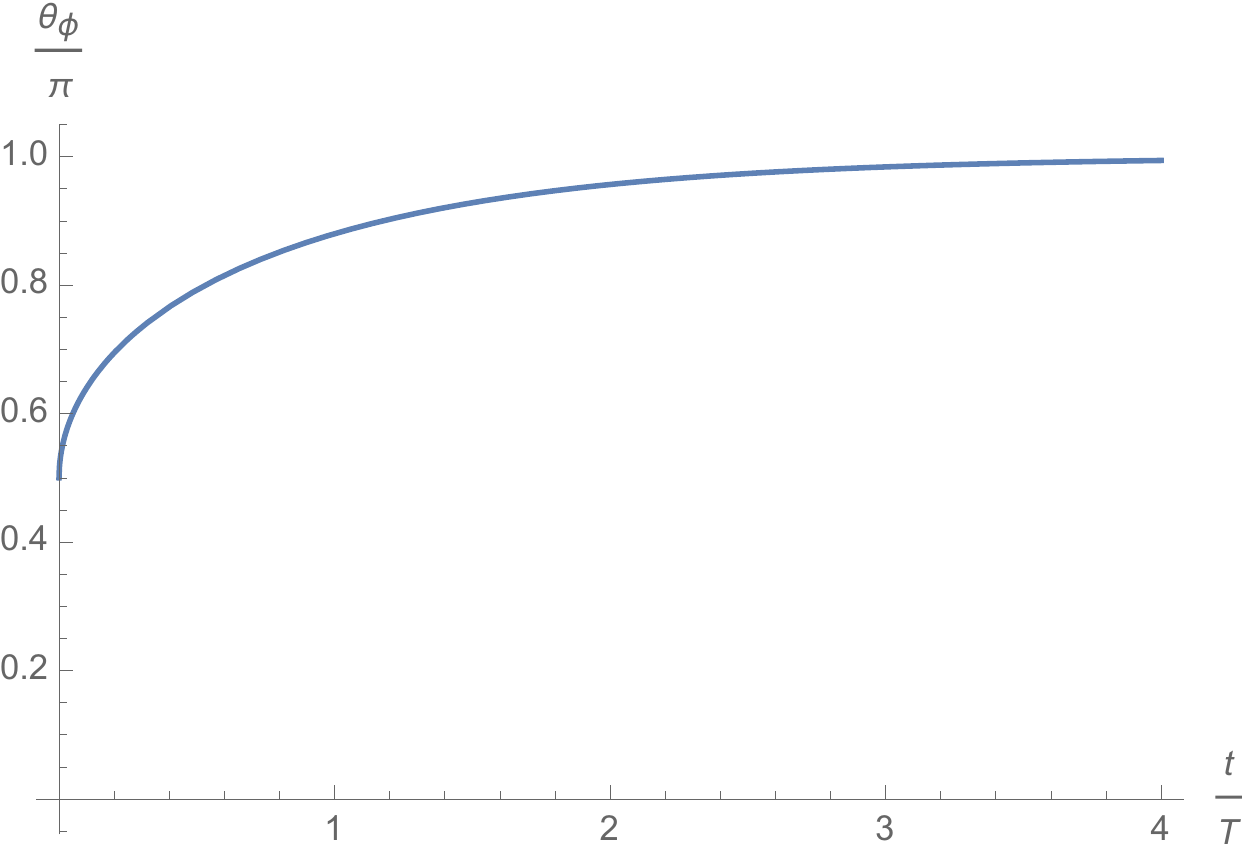}
\subcaption{Time evolution of $\theta_\varphi$}
\end{minipage}\hfill 
\begin{minipage}{.5\textwidth}
\centering \includegraphics[width=.93\linewidth]{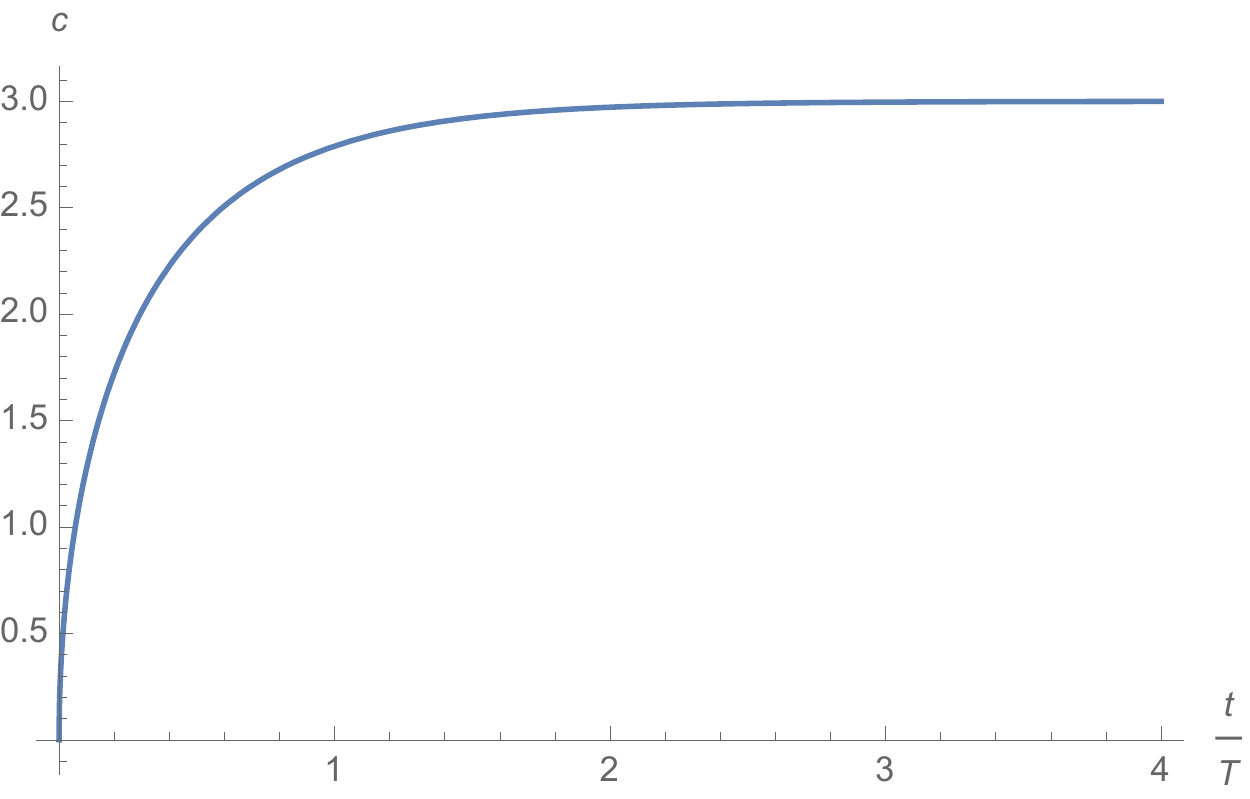}
\subcaption{Time evolution of $c$.}
\end{minipage}
\caption{{\small Approximate time evolution of $\theta_\varphi$ and
$c$ in the second order slow roll regime, where we assume that
$\theta_\varphi(0)=\frac{\pi}{2}$, i.e. $c(0)=0$. The slow
roll curve exits the rapid turn regime after the characteristic time
$T \simeq \frac{1}{\sqrt{3 V}}$ and enters the infrared regime of \cite{ren,grad}, which is
characterized by $\theta_\varphi\approx \pi$ and $c\approx 3$.}}
\label{fig:ThetacSR}
\end{figure}
Now let us make an improved estimate by taking into account the
$V_{\tau \tau}$ term as well. In that case and assuming $V, V_{\tau
\tau} \approx \const$ for simplicity, (\ref{Char_ang_Vtt}) has the
following solution:
\ben \label{th_varph_cr_AB}
\cos \theta_{\varphi} \, \approx \, \pm \,\sqrt{\frac{\hat{C} e^{2\,(A+B)\,t} + A}{\hat{C} e^{2\,(A+B)\,t} - B}} \quad ,
\een
where $\hat{C} = \const > 0$ and we defined:
\ben
A = \sqrt{3 V} \qquad , \qquad B = \frac{V_{\tau \tau}}{\sqrt{3 V}} \quad .
\een
Again this is not an exact result, since we obtained it by
assuming that $A$ and $B$ are constant. But it gives important
insights. Namely, the characteristic time is $T = \frac{1}{A+B} =
\frac{\sqrt{3V}}{3V + V_{\tau \tau}}$\,. Clearly, we can vary $T$ by
varying $V$ and $V_{\tau \tau}$\,. In any case, though, we have $\cos
\theta_{\varphi} \rightarrow \pm1$ as $t$ increases, i.e. we again have $|\sin
\theta_{\varphi}| \rightarrow 0$ for large enough $t$\,. If
$\theta_{\varphi} (0) = \frac{\pi}{2}$\,, then $\dot{\theta}_{\varphi}
> 0$ in accordance with (\ref{Th_H_c}). And if $\theta_{\varphi} (0) =
-\frac{\pi}{2}$\,, then $\dot{\theta}_{\varphi} < 0$\,. Thus in both
cases $\theta_{\varphi} (t)$ tends to $\pi \mod 2\pi$ with time, which means
that the field-space trajectory tends to align with minus the gradient
of the potential (see Figure \ref{fig:ThetacSR}). This suggests that the 
rapid turn phase is only transient, although its duration depends on the 
particulars of the potential and the field-space metric.

Since during slow roll we have $H \approx \sqrt{V/3}$\,, it follows
that the number of e-folds produced during the characteristic time (in
the above crude approximation with $V, V_{\tau \tau}\approx {\rm
const}$) is given by:
\ben
{\cal N} = \int_0^T H \,\dd t \,\approx \,\frac{V}{3 V + V_{\tau \tau}} \,\lesssim \,\frac{1}{3} \,\,\, ,
\een
where the last inequality follows from (\ref{Vtt_pos}). This suggests
that generically the rapid turn phase is rather short-lived,
likely contained within one or only a few e-folds. Obtaining a
sustained rapid turn period (capable of lasting 50 - 60 or so e-folds)
would require a rather special choice of potential.

Finally, let us comment on the quantities $f_n$ and $f_{\tau}$\,,
defined in (\ref{eta_n_eta_t}). To estimate them in the crude
approximation of (\ref{Th_vph_lead}), let us first compute the
following ratios:
\bea \label{vt_vn_r}
\frac{\dot{v}_n}{v_n} \!&=& \!\!- \eta_{\parallel} H - \dot{\theta}_{\varphi} \tan \theta_{\varphi} \approx H \left( 3 \tan^2 \theta_{\varphi} - \eta_{\parallel} \right) \approx \frac{H}{3} \,\eta_{\perp}^2 \quad , \nn \\
\frac{\dot{v}_{\tau}}{v_{\tau}} \!&=& \!\!- \eta_{\parallel} H - \dot{\theta}_{\varphi} \cot \theta_{\varphi} \approx - H \left( 3 + \eta_{\parallel} \right) \approx - 3 H \quad ,
\eea
where we used successively (\ref{dtphi_ntau}), (\ref{eta_PP}) and (\ref{Th_vph_lead}), as well as the slow roll approximation. Substituting (\ref{vt_vn_r}) in (\ref{eta_n_eta_t}) gives:
\ben
f_n \approx - \frac{\eta_{\perp}^2}{3} \qquad, \qquad f_{\tau} \approx 3 \quad .
\een
This is a rather crude estimate, relying on the leading
behavior of the characteristic angle $\theta_{\varphi} (t)$ for
$V,V_{\tau \tau} \approx {\rm const}$\,. However, it suggests that
satisfying the approximations (\ref{eta_n_eta_t_sr}) would require
a rather special choice of scalar potential.

\section{Conclusions}
\setcounter{equation}{0}
\label{sec:conclusions}

We studied consistency conditions for certain approximations commonly
used to search for rapid-turn and slow-roll inflationary trajectories
in two-field cosmological models with orientable scalar field
space. Such consistency conditions arise from requiring compatibility
between the equations of motion and the various relevant
approximations. We showed that long-lasting rapid turn trajectories
with third order slow roll can arise only in regions of field space,
where the scalar potential satisfies (\ref{VCond}). 
The latter is a nonlinear second order PDE (with coefficients depending 
on the field space metric), whose derivation follows solely from the field 
equations together with approximations (\ref{U_def_sr_rt}) and (\ref{xi_nu_cond}).
We also studied the relation between the scalar potential and scalar field
metric, which is necessary for the existence of slow-roll circular 
trajectories in two-field models with rotationally invariant scalar field 
metric. The relevant condition, given by (\ref{h-V_rel}), is solely a 
consequence of imposing the slow roll approximations $\varepsilon \ll 1$ 
and $|\eta_{\parallel}| \ll 1$ on solutions of the equations of motion 
under the assumptions of a field-space metric of the form (\ref{Grotinv}) 
and of near-circular trajectories (i.e. trajectories with $r \approx$ const). 
Finally, we showed that the approximation proposed in \cite{TB} (which is 
a rather special case of the slow-roll and rapid-turn approximation) is 
compatible with the equations of motion only when the scalar metric and 
potential satisfy a rather nontrivial consistency condition, namely 
(\ref{CCondDp}) for $\Delta \ge 0$ and (\ref{CCondDn}) for $\Delta < 0$\,, 
where $\Delta$ is the discriminant of (\ref{Cubic_pq}). The only approximations 
we used to derive the last conditions are (\ref{eta_n_eta_t_sr}) and $\varepsilon \ll 1$\,.

In principle, the various consistency conditions we have obtained could
be (approximately) satisfied numerically for a brief period by
appropriate choices of integration constants. But they can only be
maintained for a prolonged duration, if they are satisfied
functionally. Hence, given how complicated these relations are, they 
constrain dramatically the form of the scalar potential 
for a given field space metric, and vice versa. In other words, our
consistency conditions constrain severely the classes of two-field
models, for which one can hope to find long-lasting rapid-turn and
slow-roll inflationary trajectories. In that sense, our results show
that two-field cosmological models, which allow for such trajectories,
are non-generic\footnote{The intuitive notion of ``non-generic'' (equivalently, non-typical), 
actually, can be given a mathematically precise definition. In topology and algebraic geometry, a 
property is called generic if it is true on a dense open set. Further, in function space (the set 
of functions between two sets), a property is generic in ${\cal C}^n$, if it is true for a set 
containing a residual subset in the ${\cal C}^n$ topology. In our context, one can view the 
consistency conditions as maps between the set of all smooth potentials $\{V\}$ and the set of  
all smooth scalar field metrics $\{G_{IJ}\}$\,. Then the statement, that the models under 
consideration are ``non-generic'', means that the set of pairs $\{(V,G_{IJ})\}$\,, which satisfy 
our consistency conditions, has empty interior in the space of all pairs $\{(V,G_{IJ})\}$ endowed 
with (Whitney's) ${\cal C}^\infty$ topology.}
(or `rare', in the language of \cite{ACPRZ})
in the class of all two-field models. In particular, this shows that
the difficulty in finding such models -- which was previously noticed
in \cite{ACPRZ} -- is already present in the absence of supersymmetry
and hence is unrelated to supergravity.

We also studied the time evolution of the {\em characteristic angle}
of cosmological trajectories in general two-field models. We argued
that (in a crude approximation) a slow roll cosmological trajectory
with rapid turn tends to align within a short time with the opposite
of a gradient flow line of the model. Thus, generically, the solution
tends to enter the gradient flow regime of the model before producing
a sufficient number of e-folds. This confirms our conclusion that
phenomenologically viable inflationary slow roll trajectories with
long-lasting rapid turn are highly non-generic in two-field
cosmological models.\footnote{Again, by highly non-generic we mean that they 
exist only when the scalar potential has a very special form, determined by 
the relevant consistency condition, for any given scalar field metric.}

The above conclusion may be a `blessing in disguise', since finding
such rare trajectories could be more predictive (than if they were
generic) and might lead to some deep insights about the embedding of
these effective multifield models into an underlying fundamental
framework. We hope to investigate in the future what properties of the
scalar potential are implied by the consistency conditions found here,
as well as to look for ways of solving those conditions even if only
for very special choices of scalar metric. In view of \cite{BZ}, it is
also worth exploring the consequences of transient violations of slow
roll in our context.

\section*{Acknowledgements}

\noindent We thank D. Andriot, E.M. Babalic, J. Dumancic, R. Gass,
S. Paban, L.C.R. Wijewardhana and I. Zavala for interesting
discussions on inflation and cosmology. We are grateful to P. Christodoulidis 
for useful correspondence. We also thank the Stony Brook Workshop in Mathematics 
and Physics for hospitality. L.A. has received partial support from
the Bulgarian NSF grant KP-06-N38/11. The work of C.L. was supported
by grant PN 19060101/2019-2022.

\appendix

\section{Derivatives of the adapted frame} \label{DAdFr}
\setcounter{equation}{0}

\noindent Since $n$ is normalized, we have $\nabla_n (n_K n^K) = 0$\,. Hence
$\nabla_n n = n^I \nabla_I n$ does not have a component along the
vector $n$\,. To compute its component along the vector $\tau$\,, let us contract it 
with $\tau$:
\ben \label{tau_n_nabla_n}
\tau_J n^I \nabla_I n^J = \tau_J n^I \frac{G^{JK} \nabla_I V_K}{\sqrt{V_L V^L}}+\tau_J n^I G^{JK} V_K \nabla_I \!\left( \frac{1}{\sqrt{V_L V^L}} \right) \,\,\, ,
\een
where we used (\ref{ndef}) inside $\nabla_I n^J$\,. Substituting
(\ref{taudef}) in the second term of (\ref{tau_n_nabla_n}) gives:
\ben
\tau_J \nabla_n n^J = \frac{\tau^K n^I \nabla_I V_K}{\sqrt{V_L V^L}} + \sqrt{\det G} \,\epsilon_{J M} \,n^M n^I V^J \nabla_I \!\left( \frac{1}{\sqrt{V_L V^L}} \right) = \frac{V_{n \tau}}{\sqrt{V_L V^L}} \,\,\, ,
\een
where in the second equality we used the relation $\epsilon_{JM} \,n^M
V^J = \epsilon_{JM} V^M V^J / \sqrt{V_L V^L} = 0$, which follows from
(\ref{ndef}). In conclusion, we have
\ben
\nabla_n n \,= \,\frac{V_{n \tau}}{\sqrt{V_I V^I}} \,\, \tau \,\,\, .
\een
One can prove the remaining three relations in (\ref{nabla_ntau}) in a similar manner.

\section{Example: Quasi-single field inflation} \label{QSFI}
\setcounter{equation}{0}

The goal of this appendix is to demonstrate in a simple example how our consistency conditions 
can help identify the parts of field space (and/or parameter space) in which a desired type 
of inflationary solutions can exist and how to determine suitable forms of the scalar 
potential. We will not delve into the phenomenology of any particular solution, as this would take 
us too far afield from the subject of this paper. However, by studying the relevant consistency 
condition for a type of inflationary trajectories, we will show that one can derive 
constraints on the scalar field space of the model and/or find scalar potentials, which can 
support the desired inflationary regime.

For this purpose, we consider the class of models of quasi-single field inflation studied in \cite{YW}. The field space metric and scalar potential of these models are given respectively by:
\ben \label{Gf}
ds^2_G = d r^2 + f (r) d \theta^2
\een
and
\ben \label{V_quasi}
V(r, \theta) \, = \, 3 \left( W^2 (\theta) - \frac{2 W^2_{\theta} (\theta)}{3 f (r)} \right) \left[ 1 + \frac{\lambda}{2} (r - r_0)^2 + \frac{\alpha}{6} (r-r_0)^3 + ... \right]^2 \,\,\, ,
\een  
where $f(r)$ and $W(\theta)$ are arbitrary positive functions and $\alpha, \lambda, r_0 = \const$\,.\footnote{Of course, the sign of $W$ does not matter in (\ref{V_quasi}). But it will be important in relation to the Hubble parameter, as will become clear below.} In this context, one can realize orbital inflation with field-space trajectories satisfying:
\ben \label{r0}
r = r_0 \,\,\, ,
\een
and thus having $\dot{r} = 0$ identically; for more details, see \cite{YW} and references 
therein. This allows us to apply the consistency condition of Subsection \ref{SR_CC_CT}, 
which is valid for slow-rolling (near-)circular inflationary trajectories.

\vspace{0.4cm}
\noindent
$\bullet$ \hspace*{0.03cm}{\bf Slow-rolling circular trajectories:}

\vspace{0.2cm}
\noindent
To apply the consistency condition (\ref{h-V_rel}) to the above class of models, note first that 
comparing (\ref{Gf}) and (\ref{Grotinv}) implies the identification $h^2 (r) = f(r)$\,. 
Hence, in terms of $f$\,, (\ref{h-V_rel}) becomes:
\ben \label{CC_quasi}
\frac{1}{6} \frac{f_r}{f^2} \frac{V_{\theta}^2}{V_r} \approx V \,\,\, .
\een
Let us now evaluate the two sides of this relation on trajectories satisfying (\ref{r0}) in the 
above model. Substituting (\ref{V_quasi}) in (\ref{CC_quasi}), we find for the left-hand side:
\ben \label{LHS_quasi}
\frac{1}{6} \frac{f_r}{f^2} \frac{V_{\theta}^2}{V_r}\bigg|_{r = r_0} = \,3 \left( W - \frac{2}{3} \frac{W_{\theta \theta}}{f (r_0)} \right)^2 \,\,\, ,
\een
while the right-hand side gives:
\ben \label{RHS_quasi}
V|_{r = r_0} \,= \,3 \left( W^2 - \frac{2}{3} \frac{W_{\theta}^2}{f (r_0)} \right) \,\,\, .
\een

To study numerically certain phenomenological predictions of this kind of model, ref. \cite{YW} 
considered the specific choice of functions:
\ben \label{Wf_YW}
W(\theta) = C_0 \theta \qquad {\rm and} \qquad f(r) = r^2 \,\,\, ,
\een
where $C_0 = \const$\,. Using these in (\ref{LHS_quasi}) and (\ref{RHS_quasi}), 
we obtain:
\ben
\frac{1}{6} \frac{f_r}{f^2} \frac{V_{\theta}^2}{V_r}\bigg|_{r = r_0} = \,3 C_0^2 \theta^2 
\een
and
\ben \label{Vr0_YW}
V|_{r = r_0} \,= \,3 C_0^2 \theta^2 - 2 \frac{C_0^2}{r_0^2} \,\,\, .
\een
Clearly, the last two expressions are never equal. But they can be sufficiently 
numerically close to each other if the extra term in (\ref{Vr0_YW}) is small enough, 
namely if: 
\ben \label{Cond_1}
\theta^2 \gg \frac{2}{3 r_0^2} \,\,\, .
\een
Whether this condition can be satisfied or not (and for how long)
depends on the initial conditions of the trajectory under
consideration. In any case, satisfying (\ref{Cond_1}) requires a
certain level of fine-tuning of the inflationary model.\footnote{Note
that we did not use any information about the background solution for
$\theta$ in deriving (\ref{Cond_1}). In the case of approximate
solutions of the background equations of motion, the constraint
arising from the consistency condition would represent a new
approximation, in addition to slow roll. On the other hand, for exact
solutions, as those considered in \cite{YW}, satisfying the constraint
(\ref{Cond_1}) ensures slow roll, as we will show below.}

Alternatively, instead of using the ad hoc choice (\ref{Wf_YW}), we
can solve the consistency condition (\ref{CC_quasi})
functionally. This will enable us to find suitable pairs of functions
$W$ and $f$\, which are compatible with solutions of the equations of
motion in the slow roll regime. In view of (\ref{LHS_quasi}) and
(\ref{RHS_quasi}), on the trajectories of interest relation
(\ref{CC_quasi}) has the form:
\ben \label{CC_EqW}
\left( W (\theta) - \frac{2}{3} \frac{W_{\theta \theta}}{f (r_0)} \right)^2 = \,W^2 (\theta) - \frac{2}{3} \frac{W_{\theta}^2}{f (r_0)} \,\,\, .
\een
We now view (\ref{CC_EqW}) as an ODE for $W (\theta)$\,, given a function $f (r_0)$\,. The solutions 
of this equation are:
\ben \label{W_sols}
W_1 (\theta) = \frac{1}{2} C_1 \theta^2 + C_0 \theta + \frac{1}{3} \frac{C_1}{f(r_0)} + \frac{1}{2} \frac{C_0^2}{C_1} \qquad {\rm and} \qquad W_{2,3} (\theta) = e^{\pm \frac{\sqrt{6}}{2} \sqrt{f(r_0)} \,(\theta - C_{2,3})} \,\,\, ,
\een
where $C_{0,1,2,3} = \const$ and $C_1 \neq 0$\,. Note that substituting $W_{2,3} (\theta)$ in 
(\ref{RHS_quasi}) leads to $V|_{r=r_0} = 0$\,. Hence these two solutions should be discarded 
due to the requirement to have a positive potential on inflationary  
trajectories. So only $W_1 (\theta)$ is an acceptable solution in the present context. 
For later convenience, let us record its special case with $C_0 = 0$\,:
\ben \label{W_ccond}
W (\theta) = \frac{1}{2} C_1 \theta^2 + \frac{1}{3} \frac{C_1}{f(r_0)} \,\,\, .
\een

To summarize, we have shown, in principle, how the consistency conditions (in this example: for 
the existence of circular slow-roll trajectories) can either restrict the field (or parameter) 
space of an inflationary model, as in (\ref{Cond_1}), or fix the form of the potential (for a 
given field-space metric), as in (\ref{W_sols}). This illustrates manifestly, albeit on a 
technically much simpler example than in Sections \ref{sec:cons} and \ref{sec:Bjorkmo}, how our 
consistency conditions can play a constructive role in inflationary model building.

\vspace{0.4cm}
\noindent
$\bullet$ \hspace*{0.03cm}{\bf Rapid turn regime:}

\vspace{0.2cm}
\noindent
Although rapid turn inflation was not discussed in \cite{YW}, for our purposes it is interesting 
to investigate whether slow-roll circular trajectories in this kind of model can exhibit rapid 
turning in some parts of the field (and/or parameter) space. To address this question, we now 
consider the dimensionless turn rate $\eta_{\perp} = \Omega / H$ of such trajectories.

The turn rate of any background trajectory in a cosmological model with field-space metric (\ref{Gf}) is given by \cite{LA}: 
\ben \label{Om}
\Omega \, = \, \frac{\sqrt{f}}{(\dot{r}^2+f\dot{\theta}^2)} \left( \dot{\theta} V_r - \frac{\dot{r}}{f} V_{\theta} \right) \,\,\, .
\een
Also, the above quasi-single field models with potential (\ref{V_quasi}) have a Hubble parameter of the form \cite{YW}:
\ben \label{H_quasi}
H(r, \theta) \, = \, W(\theta) \left( 1 + \frac{\lambda}{2} (r - r_0)^2 + \frac{\alpha}{6} (r-r_0)^3 + ... \right) \,\,\, ,
\een
and exact inflationary solutions satisfying (\ref{r0}) and \cite{YW} 
(see also \cite{AW}):
\ben \label{th_dot}
\dot{\theta} = - 2 \,\frac{H_{\theta}}{f} \,\,\, .
\een
These solutions are not automatically slow-rolling, as will become clear below. Note also that 
(\ref{H_quasi}) implies \,$H|_{r = r_0} \!= \!W$ ; thus the requirement for positive $W$ that we 
mentioned earlier. Let us now compute the turn rate of the respective trajectories. Recall that, 
due to (\ref{r0})\,, along those trajectories $\dot{r} = 0$ identically. Hence, using 
(\ref{V_quasi}) and (\ref{Om})-(\ref{th_dot}), we find:
\ben
\Omega|_{r=r_0} \, = \, \frac{V_r}{\sqrt{f} \dot{\theta}}\bigg|_{r=r_0} = - \frac{W_{\theta} \,f_r}{f^{3/2}}\bigg|_{r=r_0}
\een
and thus
\ben \label{eta_r0}
\eta_{\perp}|_{r=r_0} = \frac{\Omega}{H}\bigg|_{r=r_0} = - \frac{W_{\theta} \,f_r}{f^{3/2} \,W}\bigg|_{r=r_0} \,\,\, .
\een

\vspace{0.4cm}
Let us now apply (\ref{eta_r0}) to the two examples discussed above. First we consider the 
choice (\ref{Wf_YW}). In that case, (\ref{eta_r0}) gives:
\ben
\eta_{\perp}|_{r=r_0} \, = \,- \,\frac{2}{r_0^2 \,\theta} \,\,\, .
\een 
Hence the rapid turn condition  $\eta_{\perp}^2 \gg 1$ implies:
\ben \label{YW_rt_c}
\frac{4}{r_0^4 \,\theta^2} \,\gg \,1 \,\,\, .
\een
Combining this with (\ref{Cond_1}), we find:\footnote{Note that using (\ref{Wf_YW}), together with (\ref{V_quasi}) and (\ref{H_quasi})-(\ref{th_dot}), gives 
\,$\epsilon_T|_{r=r_0} \!= \!\frac{1}{2f} (\frac{V_{\theta}}{V})^2|_{r=r_0} \!= \!\frac{2 \theta^2}{r_0^2 (\theta^2 - \frac{2}{3r_0^2})^2} \!\approx \!\frac{2}{r_0^2 \theta^2}$\,, where the last step is due to (\ref{Cond_1}), while \,$\varepsilon|_{r=r_0} \!= \!- \frac{\dot{H}}{H^2}|_{r=r_0} \!= \!\frac{2}{r_0^2 \theta^2}$\,; similarly, one obtains \,$\eta_T|_{r=r_0} \!= \!\frac{1}{f} \frac{V_{\theta \theta}}{V}|_{r=r_0}\!=\!\frac{2}{r_0^2 (\theta^2 - \frac{2}{3 r_0^2})} \!\approx \!\frac{2}{r_0^2 \theta^2}$ \,and \,$\eta_{\parallel}|_{r=r_0} = - \frac{\ddot{H}}{2 H \dot{H}}|_{r=r_0} = 0$\,. Hence, indeed, the constraint (\ref{Cond_1}) ensures slow roll and, in particular, the relations \,$\varepsilon \approx \epsilon_T$ \,and \,$\eta_{\parallel} \approx \eta_T - \epsilon_T$\,.}
\ben \label{C_field-sp}
\frac{2}{3} \,r_0^2 \, \ll \, r_0^4 \,\theta^2 \ll 4 \,\,\, ,
\een
which means in particular that:
\ben \label{C_par-sp}
r_0^2 \ll 6 \,\,\, .
\een
In other words, we have derived constraints on both the parameter
space of the model and the part of its field space, where the
trajectories of slow-roll rapid-turn inflationary solutions could
lie. These constraints point to different regions of field/parameter
space than those whose phenomenology was studied numerically in
\cite{YW,AW}. In these references, the numerical computations were
performed with $r_0^2 > 1$ to ensure perturbative control and to
avoid numerical instabilities (see especially \cite{AW}). Whether the
constraints (\ref{C_field-sp}) can lead to phenomenologically viable
inflationary models is an interesting question for the future.

Now let us compute (\ref{eta_r0}) for a function $W(\theta)$\,, which
solves the consistency condition (\ref{CC_EqW}). For
simplicity, we consider the special case
(\ref{W_ccond}). Substituting this $W$ in (\ref{eta_r0}) gives:
\ben
\eta_{\perp}|_{r=r_0} \, = \, - \,\frac{2 \,\theta f_r(r_0)}{f^{3/2}(r_0) \left( \theta^2 + \frac{2}{3f(r_0)} \right)} \,\,\, .
\een
Hence the rapid turn condition becomes:
\ben \label{eta_f}
\eta_{\perp}^2|_{r=r_0} \, = \, \frac{4 \,\theta^2 f_r^2}{f^{3/2} \left( \theta^2 + \frac{2}{3f} \right)^{\!2}}\Bigg|_{r=r_0} \,\gg \,1 \,\,\, .
\een
Note that so far the function $f$ has been arbitrary. This gives us a lot of freedom in how to 
satisfy (\ref{eta_f}). In particular, one can easily choose $f(r)$ such that the slow-roll and 
rapid-turn regime occurs for $r_0^2 > 1$\,, in which case one could use the same numerical methods 
as in \cite{YW,AW} to study the phenomenology of the resulting model. Clearly there is 
an unlimited number of interesting choices for $f$\,. We do not claim that any of them would be 
better than known models and/or would be compatible with further phenomenological requirements 
(nor with the third order slow roll conditions of Section \ref{sec:cons}); this is a topic for 
future studies. Our aim here was merely to illustrate in principle, and on a very simple example, 
how our consistency conditions could be a valuable guide in inflationary model building.

The discussion in this Appendix shows, furthermore, that these consistency conditions 
could provide a rich source of ideas for the development of new models or, at the very least, 
tractable toy models, which could help elucidate important features of rapid turn inflation.


\begin{thebibliography}{100}

\bibitem{GK} S. Garg, C. Krishnan, {\em Bounds on Slow Roll and the
de Sitter Swampland}, JHEP 11 (2019) 075, arXiv:1807.05193 [hep-th].

\bibitem{OPSV} H. Ooguri, E. Palti, G. Shiu, C. Vafa, {\em Distance
and de Sitter Conjectures on the Swampland}, Phys. Lett. B 788 (2019)
180, arXiv:1810.05506 [hep-th].

\bibitem{AP} A. Achucarro, G. Palma, {\em The string swampland
constraints require multi-field inflation}, JCAP 02 (2019) 041,
arXiv:1807.04390 [hep-th].

\bibitem{BPR} R. Bravo, G. A. Palma, S. Riquelme, {\em A Tip for
Landscape Riders:~Multi-Field Inflation Can Fulfill the Swampland
Distance Conjecture}, JCAP 02 (2020) 004, arXiv:1906.05772 [hep-th].

\bibitem{PT1}{C.~M.~Peterson, M.~Tegmark, {\em Testing Two-Field
Inflation}, Phys. Rev. D 83 (2011) 023522, arXiv:1005.4056
[astro-ph.CO].}

\bibitem{PT2}{C.~M.~Peterson, M.~Tegmark, {\em Non-Gaussianity in
Two-Field Inflation}, Phys. Rev. D 84 (2011) 023520, arXiv:1011.6675
[astro-ph.CO]. }

\bibitem{CAP} S. Cespedes, V. Atal, G. A. Palma, {\em On the
importance of heavy fields during inflation}, JCAP 05 (2012) 008,
arXiv:1201.4848 [hep-th].

\bibitem{BFM} T. Bjorkmo, R. Z. Ferreira, M.C.D. Marsh, {\em Mild
Non-Gaussianities under Perturbative Control from Rapid-Turn Inflation
Models}, JCAP 12 (2019) 036, arXiv:1908.11316 [hep-th].

\bibitem{PSZ} G. A. Palma, S. Sypsas, C. Zenteno, {\em Seeding
primordial black holes in multi-field inflation}, Phys. Rev. Lett. 125
(2020) 121301, arXiv:2004.06106 [astro-ph.CO].

\bibitem{FRPRW} J. Fumagalli, S. Renaux-Petel, J. W. Ronayne,
L. T. Witkowski, {\em Turning in the landscape:~a new mechanism for
generating Primordial Black Holes}, arXiv:2004.08369 [hep-th].

\bibitem{LA} L. Anguelova, {\em On Primordial Black Holes from Rapid
Turns in Two-field Models}, JCAP 06 (2021) 004, arXiv:2012.03705
[hep-th].

\bibitem{LA2} L. Anguelova, {\em Primordial Black Hole Generation in a 
Two-field Inflationary Model}, Springer Proc. Math. Stat. 396 (2022) 193,  
arXiv:2112.07614 [hep-th].

\bibitem{ASSV} Y. Akrami, M. Sasaki, A. Solomon and V. Vardanyan, {\em
Multi-field dark energy: cosmic acceleration on a steep potential},
Phys. Lett. B 819 (2021) 136427, arXiv:2008.13660 [astro-ph.CO].

\bibitem{ADGW} L. Anguelova, J. Dumancic, R. Gass,
L.C.R. Wijewardhana, {\em Dark Energy from Inspiraling in Field
Space}, JCAP 03 (2022) 018, arXiv:2111.12136 [hep-th].

\bibitem{AB} A Brown, {\em Hyperinflation}, Phys. Rev. Lett. 121 (2018)
251601, arXiv:1705.03023 [hep-th].

\bibitem{SM}
S. Mizuno, S. Mukohyama, {\em Primordial perturbations from inflation with 
a hyperbolic field-space}, Phys. Rev. D 96 (2017) 103533, arXiv:1707.05125 [hep-th].

\bibitem{CRS} P. Christodoulidis, D. Roest, E. Sfakianakis, {\em
Angular inflation in multi-field $\alpha$-attractors}, JCAP 11 (2019)
002, arXiv:1803.09841 [hep-th].

\bibitem{GSRPR} S. Garcia-Saenz, S. Renaux-Petel, J. Ronayne, {\em
Primordial fluctuations and non-Gaussianities in sidetracked
inflation}, JCAP 1807 (2018) 057, arXiv:1804.11279 [astro-ph.CO].

\bibitem{ACIPWW} A. Achucarro, E. Copeland, O. Iarygina, G. Palma,
D.G. Wang, Y. Welling, {\em Shift-Symmetric Orbital Inflation:~single
field or multi-field?}, Phys. Rev. D 102 (2020) 021302,
arXiv:1901.03657 [astro-ph.CO].

\bibitem{TB} T. Bjorkmo, {\em The rapid-turn inflationary attractor},
Phys. Rev. Lett. 122 (2019) 251301, arXiv:1902.10529 [hep-th].

\bibitem{BM} T. Bjorkmo, M. C. D. Marsh, {\em Hyperinflation generalised: 
from its attractor mechanism to its tension with the `swampland conditions'},
JHEP 04 (2019) 172, arXiv:1901.08603 [hep-th].

\bibitem{APR}
V. Aragam, S. Paban, R. Rosati, {\em The Multi-Field, Rapid-Turn Inflationary Solution}, 
JHEP 03 (2021) 009, arXiv:2010.15933 [hep-th].

\bibitem{ZGZ}
Y. Zhang, Y.-g. Gong, Z.-H. Zhu, {\em Noether Symmetry Approach in multiple scalar fields 
Scenario}, Phys. Lett. B688 (2010) 13, arXiv:0912.0067 [hep-ph].

\bibitem{PT}
A. Paliathanasis, M. Tsamparlis, {\em Two scalar field cosmology: Conservation laws and 
exact solutions}, Phys. Rev. D 90 (2014) 043529, arXiv:1408.1798 [gr-qc].

\bibitem{ABL}
L. Anguelova, E.M. Babalic, C.I. Lazaroiu, {\em Two-field Cosmological
$\alpha$-attractors with Noether Symmetry}, JHEP 1904 (2019) 148,
arXiv:1809.10563 [hep-th].

\bibitem{ABL2} L. Anguelova, E.M. Babalic, C.I. Lazaroiu, {\em Hidden
symmetries of two-field cosmological models}, JHEP 09 (2019) 007,
arXiv:1905.01611 [hep-th].

\bibitem{Hesse} C. I. Lazaroiu, {\em Hesse manifolds and Hessian
symmetries of multifield cosmological models}, Rev. Roumaine
Math. Pures Appl. 66 (2021) 2, 329-345, arXiv:2009.05117 [hep-th].

\bibitem{genalpha}{C. I. Lazaroiu, C. S. Shahbazi, {\em Generalized
two-field $\alpha$-attractor models from geometrically finite
hyperbolic surfaces}, Nucl. Phys. B 936 (2018) 542-596, arXiv:1702.06484 [hep-th].}

\bibitem{elem}{E. M. Babalic, C. I. Lazaroiu, {\em Generalized
$\alpha$-attractor models from elementary hyperbolic surfaces},
Adv. Math. Phys. 2018 (2018) 7323090, arXiv:1703.01650 [hep-th].}

\bibitem{modular}{E. M. Babalic, C. I. Lazaroiu, {\em Generalized
$\alpha$-attractors from the hyperbolic triply-punctured sphere},
Nucl. Phys.  B 937 (2018) 434-477, arXiv:1703.06033 [hep-th].}

\bibitem{ren}{C. I. Lazaroiu, {\em Dynamical renormalization and
universality in classical multifield cosmological models},
Nucl. Phys. B 983 (2022) 115940, arXiv:2202.13466 [hep-th].}

\bibitem{grad} E. M. Babalic, C. I. Lazaroiu, {\em The infrared
behavior of tame two-field cosmological models}, Nucl. Phys. B 983
(2022) 115929, arXiv:2203.02297 [gr-qc].

\bibitem{ACPRZ} V. Aragam, R. Chiovoloni, S. Paban, R. Rosati,
I. Zavala, {\em Rapid-turn inflation in supergravity is rare and
tachyonic}, JCAP 03 (2022) 002, arXiv:2110.05516 [hep-th].

\bibitem{LPB}{A, R. Liddle, P. Parsons, J. D. Barrow, {\em Formalizing
the slow-roll approximation in inflation}, Phys. Rev. D 50 (1994)
7222, arXiv:astro-ph/9408015.}

\bibitem{HP} A. Hetz, G. Palma, {\em Sound Speed of Primordial
Fluctuations in Supergravity Inflation}, Phys. Rev. Lett. 117 (2016)
101301, arXiv:1601.05457 [hep-th].

\bibitem{AGHPP} A. Achucarro, J.-O. Gong, S. Hardeman, G. Palma,
S. Patil, {\em Features of heavy physics in the CMB power spectrum},
JCAP 01 (2011) 030, arXiv:1010.3693 [hep-ph].

\bibitem{CCLBNZ} D. Chakraborty, R. Chiovoloni, O. Loaiza-Brito,
G. Niz, I. Zavala, {\em Fat inflatons, large turns and the
$\eta$-problem}, JCAP 01 (2020) 020, arXiv:1908.09797 [hep-th].

\bibitem{CRS3}
P. Christodoulidis, D. Roest, E. Sfakianakis, {\em Attractors, Bifurcations 
and Curvature in Multi-field Inflation}, JCAP 08 (2020) 006, arXiv:1903.03513 [gr-qc].

\bibitem{CRS2}
P. Christodoulidis, D. Roest, E. Sfakianakis, {\em Scaling attractors in 
multi-field inflation}, JCAP 12 (2019) 059, arXiv:1903.06116 [hep-th].

\bibitem{BZ} S. Bhattacharya, I. Zavala, {\em Sharp turns in axion
monodromy: primordial black holes and gravitational waves},
arXiv:2205.06065 [astro-ph.CO].

\bibitem{YW}
Y. Welling, {\em A simple, exact, model of quasi-single field inflation}, 
Phys. Rev. D 101, 063535 (2020), arXiv:1907.02951 [astro-ph.CO].

\bibitem{AW}
A. Achucarro, Y. Welling, {\em Orbital Inflation: inflating along an angular isometry 
of field space}, arXiv:1907.02020 [hep-th].

\end{thebibliography}
\end{document}